\documentclass[final,5p,times,twocolumn]{elsarticle}

\usepackage{overpic}
\usepackage{amssymb}

\newcommand{\cross}{\nabla\times}
\newcommand{\micron}{$~\mathrm{\mu}$m~}
\newcommand{\degree}{\ensuremath{^\circ}}

\journal{High Energy Density Physics}

\begin{document}

\begin{frontmatter}



\title{Prospects of Turbulence Studies in High-Energy Density Laser-Generated Plasma:\\Numerical Investigations in Two Dimensions}


\author[fsu]  {Timothy Handy\corref{cor1}}
\ead{tah09e@fsu.edu}
\author[fsu]  {Tomasz Plewa}
\ead{tplewa@fsu.edu}
\author[umich]{R. Paul Drake}
\ead{rpdrake@umich.edu}
\author[kit]{Andrey Zhiglo}
\ead{azhiglo@gmail.com}

\cortext[cor1]{Corresponding Author: Florida State University,
  Department of Scientific Computing, 400 Dirac Science Library,
  Tallahassee, FL 32306-4120, USA. Phone: (850) 644-1010 Fax: (850)
  644-0098, Email: tah09e@fsu.edu}

\address[fsu]{Department of Scientific Computing\\Florida State University\\Tallahassee, FL, USA}

\address[umich]{Center for Radiative Shock Hydrodynamics\\Department
  of Atmospheric, Oceanic and Space Sciences\\University of
  Michigan\\Ann Arbor, MI, USA}

\address[kit]{NSC Kharkov Institute of Physics and Technology\\Kharkov, Ukraine}
\begin{abstract}

We investigate the possibility of generating and studying turbulence in plasma by means of high-energy density laser-driven experiments. Our focus is to create supersonic, self-magnetized turbulence with characteristics that resemble those found in the interstellar medium (ISM). 

We consider a target made of a spherical core surrounded by a shell made of denser material. The shell is irradiated by a sequence of laser pulses sending inward-propagating shocks that convert the inner core into plasma and create turbulence. In the context of the evolution of the ISM, the shocks play the role of supernova remnant shocks and the core represents the ionized interstellar medium. We consider the effects of both pre-existing and self-generating magnetic fields and study the evolution of the system by means of two-dimensional numerical simulations.

We find that the evolution of the turbulent core is generally, subsonic with rms-Mach number $M_t\approx 0.2$. We observe an isotropic, turbulent velocity field with an inertial range power spectra of $P(k)\propto k^{-2.3}$. We account for the effects of self-magnetization and find that the resulting magnetic field has characteristic strength $\approx 3\times10^{4}$ G. The corresponding plasma $\beta$ is about $1\times10^{4}-1\times10^{5}$, indicating that the magnetic field does not play an important role in the dynamical evolution of the system.

The natural extension of this work is to study the system evolution in three-dimensions, with various laser drive configurations, and targets with shells and cores of different masses. The latter modification may help to increase the turbulent intensity and possibly create transonic turbulence. One of the key challenges is to obtain transonic turbulent conditions in a quasi-steady state environment. 

\end{abstract}
\begin{keyword}
hydrodynamics \sep turbulence \sep  magnetic fields \sep laboratory astrophysics
\end{keyword}

\end{frontmatter}
%
%
\label{s:intro}

The main physics processes describing the evolution of the interstellar medium (ISM) are hydrodynamics, magnetization, and radiation processes such as ionization. In addition, the stars are localized sources of mass and energy, some of them, such as supernovae, are very powerful and capable of shaping the evolution and structure of the ISM on global scales. Those interactions eventually result in small-scale structures, including supersonic turbulence.

Absorption and emission at infrared and radio wavelengths are the primary messengers for turbulence in the interstellar medium (ISM). Observations have shown that velocity dispersion is correlated to region size via a power law dependence from sub-parsec to kilo-parsec scales \cite{larson+79,larson+81,ossenkopf+02}. Observations of velocity and density are consistent with supersonic turbulence driven on large scales (at or above the size of molecular clouds), and exhibit velocity structures indicative of a shock-dominated medium \cite{ossenkopf+02}. There is further evidence that small-scale driving from star formation is negligible \cite{brunt+02a,brunt+02b,brunt+09}, and the observational velocity scaling is inertial. Radio scintillation measurements provide direct evidence for turbulence on small-scales ($\sim 10^{12}$~cm) \cite{rickett+90,boldyrev+03}.

The existence of supersonic (compressible) turbulence plays an important role in star formation \cite{maclow+04,krumholz+05,padoan+11}, the stellar initial mass function \cite{padoan+02,hennebelle+08,hennebelle+09}, and more fundamentally, the density and velocity statistics of the ISM \cite{elmegreen+04,scalo+04,mckee+07}. The primary metric in these analyzes is the dependence on the density variance with the rms Mach number; for log-normal distributions of the density contrast ($s\equiv \ln{\left(\rho/\rho_0\right)}$), the density variance is given by $\sigma_s = \ln{\left(1+b^2M^2\right)}$, where $b$ is related to the energy injection mechanism \cite{kritsuk+07,lemaster+08,padoan+11,molina+12}. Despite analytical \cite{padoan+11, molina+12} and numerical \cite{ostriker+01,lemaster+08} investigations, the effect of magnetic fields on the density variance remains unclear.

Information about the magnetic field in the ISM can be inferred via Faraday rotation and polarization of synchrotron radiation \cite{troland+08,crutcher+09}. Observations based on Zeeman measurements indicate that magnetic effects may play an important role in hydrodynamic evolution \cite{crutcher+99}. However, the sustenance of the interstellar magnetic field may be be coupled to turbulence via generation due to folding and stretching of the field, resulting in small-scale dynamo effects \cite{schekochihin+04}.

Earth-bound methods for investigating turbulence in the ISM have come only from numerical modeling. Numerical simulations indicate that the webbed structure of the ISM may result from the nonlinear advection operator \cite{scalo+98}. Additionally, pure hydrodynamic simulations reproduce similar behavior \cite{kritsuk+07}. Marginal quantitative differences are found between hydrodynamic and magnetohydrodynamic simulations when subjected to a background magnetic field \cite{deavillez+05}; however, filaments tend to orient along magnetic field lines. There is numerical evidence that energy transfer between spatial scales may be regulated by shocks, as opposed to a turbulent cascade \cite{lemaster+09}.

While turbulence has been studied extensively in fluids \cite{frisch+95,kellay+02,boffetta+12}, it has not received the same treatment for plasmas. However, advancements in high-energy-density physics (HEDP) experiments may provide a way to investigate properties of the ISM in the laboratory. Laser systems at the Omega Laser Facility and the National Ignition Facility provide the ability to deposit kilojoule to megajoule energies on the surface of millimeter scale targets over a timescale of picoseconds to nanoseconds. Experimental approaches to potential turbulence inducing effects from Rayleigh-Taylor, Kelvin-Helmholtz, and Richtmyer-Meshkov instabilities have been summarized in \cite{drake+08a}. Replication of turbulence in the ISM will require long driving times, implying the need to drive large amounts of material (large volumes) presumably using multiple blast wave-like impulses created by an array of laser beams. 

An initial experimental design to study compressible turbulence in the laboratory was proposed by \cite{drake+08b}. In this design, a gas-filled target box is embedded in a medium. As this surrounding medium is exposed to laser energy deposition, small holes in the box allow driven material to enter the cavity. The interactions of these focused blast waves are then expected to produce turbulent behavior in the interior of the container. 

In this work we present a preliminary high-energy-density experimental scenario to investigate shock generated turbulence. Our initial, two-dimensional scenario focuses on stirring a target composed of concentric, circular layers with blast waves generated from laser irradiation. We include the effects of magnetic fields in this work by considering three cases; pure hydrodynamics, a pre-existing magnetic field, and a magnetic field generated from the Biermann battery source term. Section~\ref{s:methods} outlines our experimental design and the corresponding computational model, with supplementary material found in \ref{s:crashlaser}. Section~\ref{s:results} provides a comparison between the three magnetization cases. Section~\ref{s:selfgenerated} offers an in-depth analysis of the self-generated magnetic field case. Discussion and conclusions are offered in Section~\ref{s:conc}
\section{Design, model, \& methods}
\label{s:methods}

%
%
%
%

\subsection{Experimental scenario}
\label{s:proposal}
\begin{figure*}[ht!]
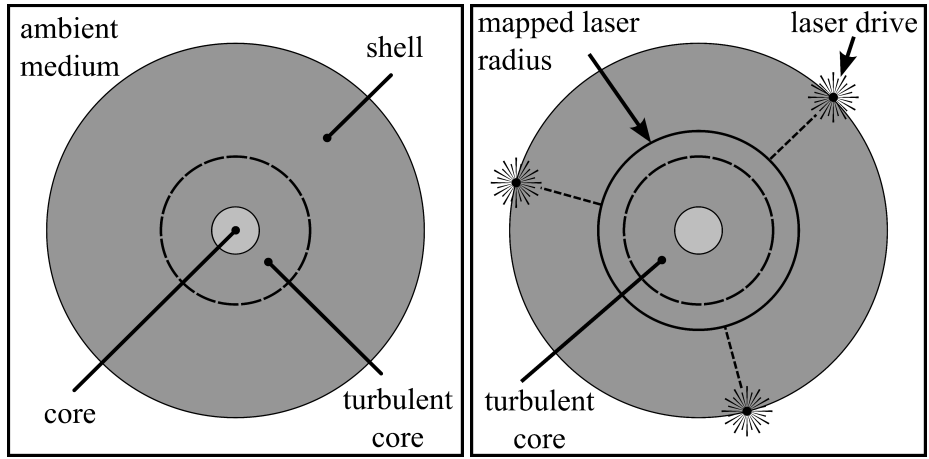

	\begin{center}
		\includegraphics[width=6cm]{f1_L.pdf}
		\includegraphics[width=6cm]{f1_R.pdf}
	\end{center}
\caption{Proposed experimental setup. (left panel) Two-material target configuration. The dashed circle shows the region we consider for our analysis, the ``turbulent core''. (right panel) Schematic for one round of laser driving. Three lasers (a triple) are positioned 120\degree apart. Each laser is a precomputed hydrodynamic profile from the CRASH code, with the leading edge of the shock touching the thick solid line.}
\label{f:diagram}
\end{figure*}

Our preliminary experimental design is based on two concentric spheres as illustrated in Fig.~\ref{f:diagram}. The inner sphere (core) is composed of low density material and provides a medium for the driven mixing process. The outer sphere (shell) is of higher density. The target is embedded in a very low density ambient medium (essentially vacuum). All materials are initially in pressure equilibrium. The aim of this work is to produce an initial investigation of the two-dimensional problem.

In order to reproduce a multiply shocked section of interstellar medium, the shell layer is irradiated on its surface by a set of laser drives. The drive-shell interactions result in pressure and material waves propagating toward the center of the target. As the shocks pass through the core at different times complex hydrodynamic conditions are created. The primary purpose of the high density shell is to absorb the laser drive and convert thermal energy into kinetic energy. As a result, the perturbations reaching the core region should be  kinetically dominated. Our hope is repeated exposure of the shell material to laser pulses should produce a proxy for the effects of supernovae in the ISM. Ideally, the laser system would be arranged in a spherically symmetric configuration to help confinement of the target material and allow for longer evolution of the shocked system. 

The effects of laser driving on the shell material is a transient problem that is not of particular interest in this work. We are primarily interested in the effects of mixing in the light core, and therefore consider a ``turbulent core'' (TC) encompassing the target core and part of the shell. Consequently, we view the area exterior to the TC as generating the boundary conditions for the interior of the TC. The results of this work focus only on data interior to the TC, unless otherwise stated.

\subsection{Computational model}

We assume that the laboratory setting can be modeled via the extended magnetohydrodynamic equations outlined in \cite{braginskii+65,nishiguchi+02}:
\begin{equation}
	\label{e:mhd-cnt}
	\frac{\partial\rho}{\partial t} + \nabla\cdot\left(\rho\bf{u}\right) = 0,
\end{equation}
\begin{equation}
	\label{e:mhd-mom}
	\frac{\partial\rho\bf{u}}{\partial t} + \nabla\cdot\left(\rho\bf{u}\otimes\bf{u}\right) = -\nabla\left(p+\frac{\left|\bf{B}\right|^2}{8\pi}\right),
\end{equation}
\begin{equation}
	\label{e:mhd-nrg}
	\frac{\partial\rho\varepsilon}{\partial t} + \nabla\cdot\left(\rho\varepsilon\bf{u}\right) = -\nabla\cdot\left(p\bf{u}\right),
\end{equation}
\begin{equation}
	\label{e:mhd-mag}
	\frac{\partial\bf{B}}{\partial t} = \cross\left(\bf{u}\times\bf{B}\right) + \frac{c}{e}\left[\cross\frac{\nabla p_e}{n_e} - \cross\frac{\left(\cross\bf{B}\right)\times\bf{B}}{4\pi n_e}\right],
\end{equation}
where $\rho$, $p$, $\varepsilon$, $\bf{u}$, $\bf{B}$ are the fluid density, thermal pressure, specific total energy, material velocity, and magnetic field, respectively. The electron pressure and electron density are denoted by $p_e$ and $n_e$, respectively. The leading constants on the bracketed term of Eqn.~\ref{e:mhd-mag} are the speed of light ($c$) and electron charge ($e$).

Equation~\ref{e:mhd-mag} describes the evolution of the magnetic field using the generalized Ohm's law, where the bracketed term indicates the components that result in self-generation of the magnetic field. The first term inside the brackets is the Biermann battery term, and causes field generation when the temperature and density gradients are misaligned. The second term is the Hall term, and indicates that pre-existing fields can self amplify. When no initial magnetic fields exist, the Biermann battery term may begin creating a field, which the Hall term will then act upon. We do not include the electron-ion friction term in our models.

We utilize an ideal gas equation of state with $\gamma = 1.6$. The plasma composition is represented with a single species with atomic mass, $A$, and atomic charge, $Z$. Accordingly, the ion number density is $n_i = \rho N_\mathrm{A}/A$, where $N_\mathrm{A}$ is the Avogadro constant. To obtain the electron number density, $n_e$, we use the Thomas-Fermi equation of state \cite{salzmann+98}. The required electron number density is calculated as $n_e={\bar Z}n_i$.

\subsection{Proteus}

The set of Eqns.~(\ref{e:mhd-cnt})-(\ref{e:mhd-mag}) are solved numerically using the finite volume Proteus code, which is our developmental fork of the FLASH code \cite{fryxell+00}. In this work we use the unsplit staggered mesh solver of \cite{lee+09}. This magneto-hydrodynamics (MHD) solver is a variant of the constrained transport method \cite{evans+88} and is used for all cases considered. The MHD solver is formally second-order accurate in space and time.

The method used for driving the turbulence in these models (Sec.~\ref{s:laser}) produces numerical difficulties for many Riemann solvers. In order to maintain a robust simulation environment, we use the Harten-Lax-van Leer-Einfeldt (HLLE) Riemann solver \cite{einfeldt+88}.

Computations are performed on a square, Cartesian domain with sides of length 9000\micron. This allows for the entire target to be placed inside, with additional room to develop in the ambient medium. We treat the boundary conditions as open outflows. While we limit our analysis to the smaller region of interest surrounding the core, simulation of the complete target allows for the interaction of laser-driven material. As the area exterior to the TC can be viewed as boundary conditions for the TC, it is possible that these interactions could affect our results and are accounted for as much as possible. 

We utilize statically refined Cartesian meshes for our domain decomposition. We refine the grid inside of the region of interest to a uniform spacing. We perform each case on three separate TC grid sizes: 16\micron~for the coarsest run; 8\micron~for the medium resolution run; and 4\micron~for the fine resolution run. Exterior to the TC we allow the mesh to coarsen radially (subjected to proper nesting). The coarsest cell resolution for all computed models is 64\micron.

\subsection{Initial conditions}
\label{s:expcond}

\begin{table}[h]
	\centering
	\caption{Target geometry and initial conditions.}
	\begin{tabular}{|c|r l|}
	\hline
	$\rho_{ambient}$&$1\times10^{-4}$&g$~$cm$^{-3}$\\ \hline
	$\rho_{shell}$&$1\times10^{-1}$&g$~$cm$^{-3}$\\ \hline
	$\rho_{core}$&$2\times10^{-2}$&g$~$cm$^{-3}$\\ \hline
	$p_{target}$&$1\times10^{3}$&bar\\ \hline
	$r_{shell}$&$3250$&\micron\\ \hline
	$r_{core}$&$500$&\micron\\ \hline
	$r_{TC}$&$1500$&\micron\\ \hline
	$T_{core}$&$11,600$&K\\ \hline
	\end{tabular}
	\label{t:ic_table}
\end{table}

We consider three distinct magnetization cases in this work: pure hydrodynamics; a pre-existing, out of plane magnetic field; and self-generated magnetic fields (no pre-exiting field). The pure hydrodynamics case provides a reference from which to gauge the impact of magnetic fields on the flow structure. With these three cases, we aim to begin quantifying the effects of turbulence in high energy density conditions, attempting to include increasing levels of physical complexity. Table~\ref{t:ic_table} shows the shared conditions for all of our models.

The pre-existing magnetic field case consists of an initial magnetic field whose only nonzero component is out of the plane. This field provides an additional pressure component (in the form of magnetic pressure), without forcing the advection of material to be along field lines. Such a configuration is difficult to realize in the laboratory, and should be thought of as a toy model. However, it allows us to gauge any magnetic effects in a controlled manner, where we have an \emph{a priori} estimate of the plasma $\beta$. (For the entirety of this work we define $\beta$ as $\beta=p/(8\pi B^2)$.) In order to initialize the magnetic field we use a characteristic reference pressure of $p_{ref}=1\times10^{6}$ bar and choose $\beta_{ref}=20$ to solve for the magnitude of the initial magnetic field. Our chosen reference pressure is characteristic of conditions in the turbulent core during the laser driving period, and should put our pre-existing field simulations near $\beta_{ref}$. 

\subsection{Shock generation}
\label{s:laser}

The requirement of producing high-energy density turbulent plasma in a state
close to steady state and as isotropic as possible imposes certain
restrictions on the laser drive. In a viable design it would be highly
desirable that average linear and angular momenta of the system are close to
zero. One possible way to achieve this is to compensate for the linear
momentum injected by any laser pulse using a some combination of remaining
laser pulses. For example, one could fire two laser beams from opposite
directions. This configuration, however, would not allow for lasting plasma
confinement. Therefore, a more complex setting is needed, such as a triple
(with laser beams originating at the tips of an equilateral triangle in 2D) or
a quadruple (with laser beams originating at the tips of a tetrahedron in 3D)
laser drive configuration. Furthermore, one would wish to add a certain degree
of randomness to the drive in order to promote the development of turbulence.
(Although such a drive configuration cannot be realized at existing HEDP laser
facilities, our primary goal is to assess the feasibility of possible future
designs for turbulent plasma experiments.)

In our two-dimensional study, the laser drive configuration is defined by a
set of three lasers arranged 120$^\circ$ apart (a triple) in order to improve
confinement. Over the course of driving we sample 100 triples every 2
nanoseconds with each laser in the triple fired at the same time. Each triple
is offset by a random angle, $\theta$, sampled from a uniform distribution
over $0^\circ<\theta<120^\circ$. The initial triplet has an angular offset of
$\sqrt{2}\times180^\circ/\pi\approx81^\circ$ in order to avoid initial grid
symmetries.

While we use Proteus to evolve the target, we do not use it to simulate the laser-target interaction. Instead, we precompute a two-dimensional ``laser drive profile'' (LDP) and then map it onto the Proteus mesh when a triple is activated. The LDP is a fixed time, two dimensional, cylindrical set of primitive variables, and is further described in \ref{s:crashlaser}. When mapping the LDP to the Proteus mesh, we place the inward moving tip along a circle of radius $r_{map}=2000$\micron, with the angle of the LDP axis given by the sampled angle $\theta$. 

We note that the process of mapping the LDP onto the Proteus mesh is not conservative. We feel this approach is justified, as the drive is mapped away from the TC, which is an open system in its own right. Thus, the area outside is of ancillary importance in terms of conservation. This method also enables dramatically faster turnaround on model generation, as we do not have to compute laser-target interactions in the complex media surrounding the TC. More realistic studies aimed at evaluating specific laboratory experiments may have to abandon the LDP concept and compute the laser energy deposition along with the interior calculations at the significant increase of computational time. Again, our aim in this work is to provide initial insights into the behavior of such an experiment.
\section{Effect of magnetic fields on hydrodynamics}
\label{s:results}

\subsection{Compressibility effects}
\begin{figure}[h!]
	\centering
	\includegraphics[width=8cm]{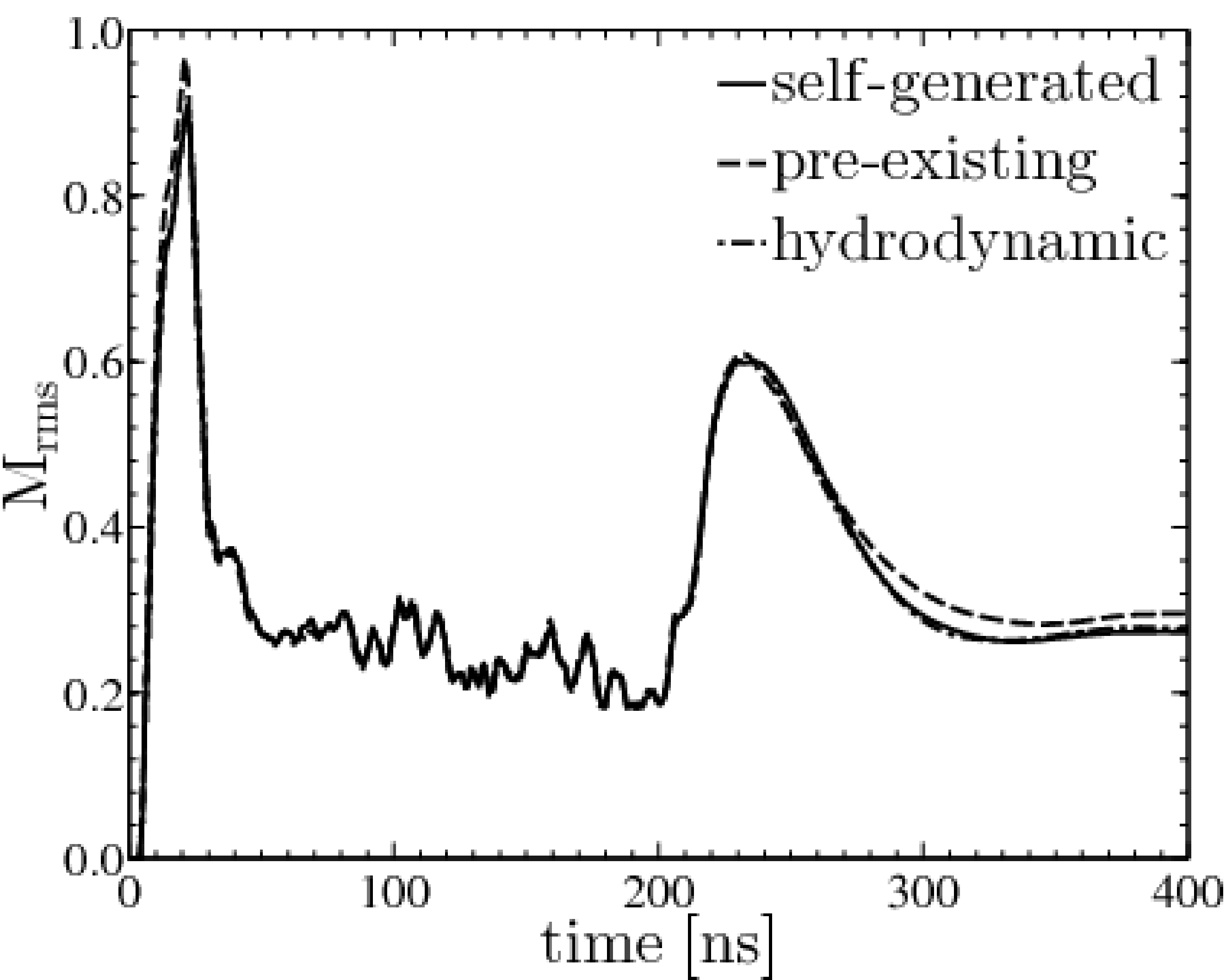}
	\caption{RMS-Mach number for the hydrodynamic, pre-existing magnetic field, and self-generating magnetic field cases. The Mach number behaves similarly for all magnetization cases, indicating that the magnetic field is likely too low to influence the hydrodynamic development of the system. The laser-driven, quasi-steady state system never reaches the supersonic regime.}
\label{f:2d-mach_turb}
\end{figure}
In order to judge the effects of compressibility, we consider the evolution of the rms-Mach number, $M_{rms}$, shown in Fig.~\ref{f:2d-mach_turb} for the hydrodynamic, pre-existing field, and self-generated cases. The first laser-driven shocks penetrate the low density core at $t=30$~ns, causing a sharp rise in $M_{rms}$. By $t=50$~ns the core has been completely overrun and the rms-Mach number reaches a quasi-steady value of approximately 0.2. 

For the remaining $t=150$~ns of evolution, continual driving via laser
triplets stirs the region of interest. Over this period the rms-Mach number
marginally decays due to the combined effects of suppressed material
accelerations (resulting from the confinement via the laser arrangement) and
sound speed increase via compression. During this driving phase there is no
discernible effect from either \emph{a priori} or \emph{in situ} magnetic
fields. The low rms-Mach numbers obtained during the driving phase indicates
that we are not reaching the supersonic regime. When laser driving ceases at
$t=200$~ns there is a marked increase in the rms- Mach number. This trend
peaks at $t\approx230$~ns, after which $M_{rms}$ decays to a nominal value of
0.25.

After the laser drive turns off, there is no confining ram pressure to balance the
thermal pressure in the core. This results in outward expansion which
simultaneously increases material velocity and decreases sound speed. The Mach
number is able to stabilize at later times as conditions in the region of
interest homogenize. We note that the deviation between $M_{rms}$ and the
density-weighted turbulence Mach number is approximately 3\% during the
driving phase.

\subsection{Magnetic field evolution}
\begin{figure}[h!]

	\centering
	\includegraphics[width=8cm]{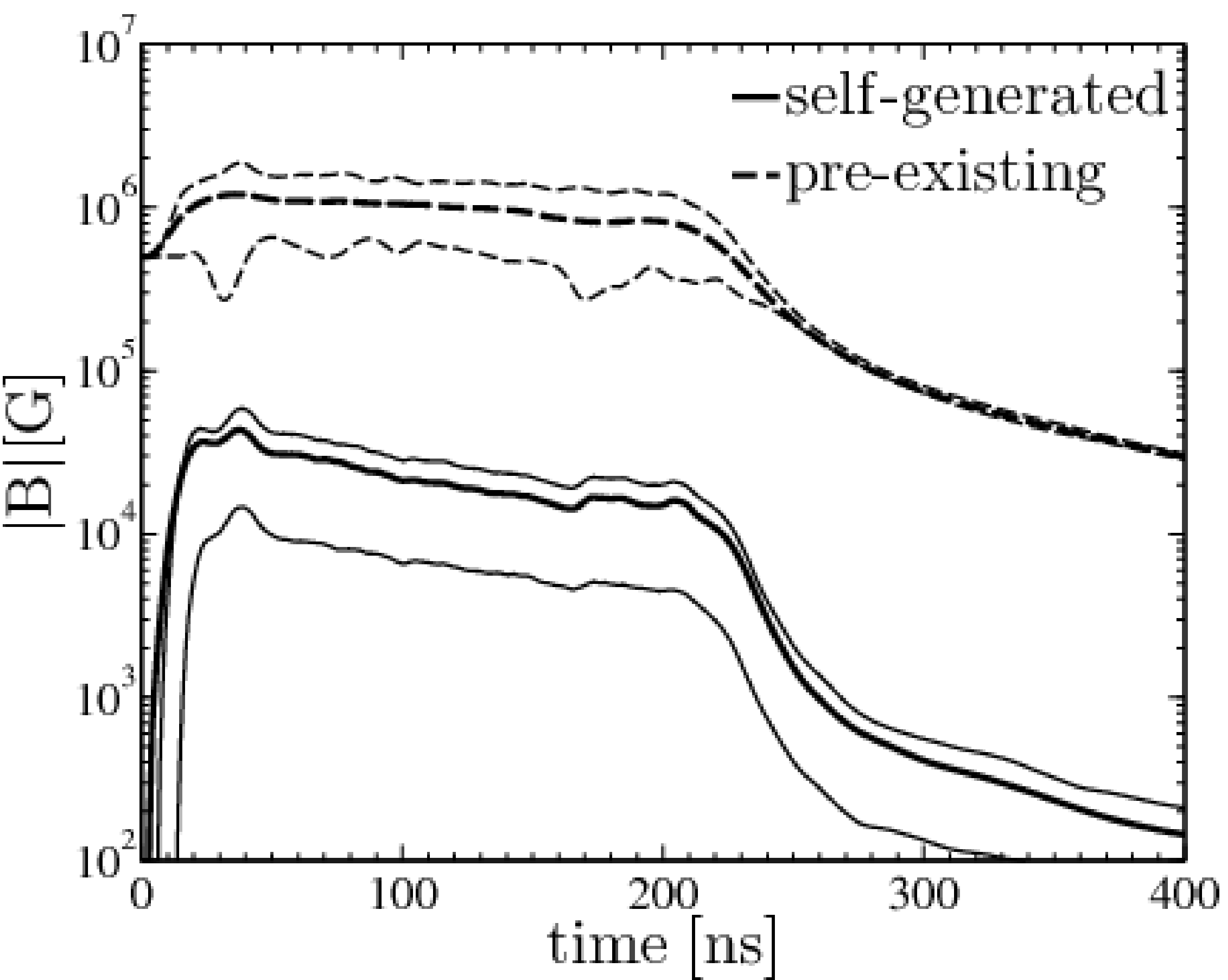}
	\includegraphics[width=8cm]{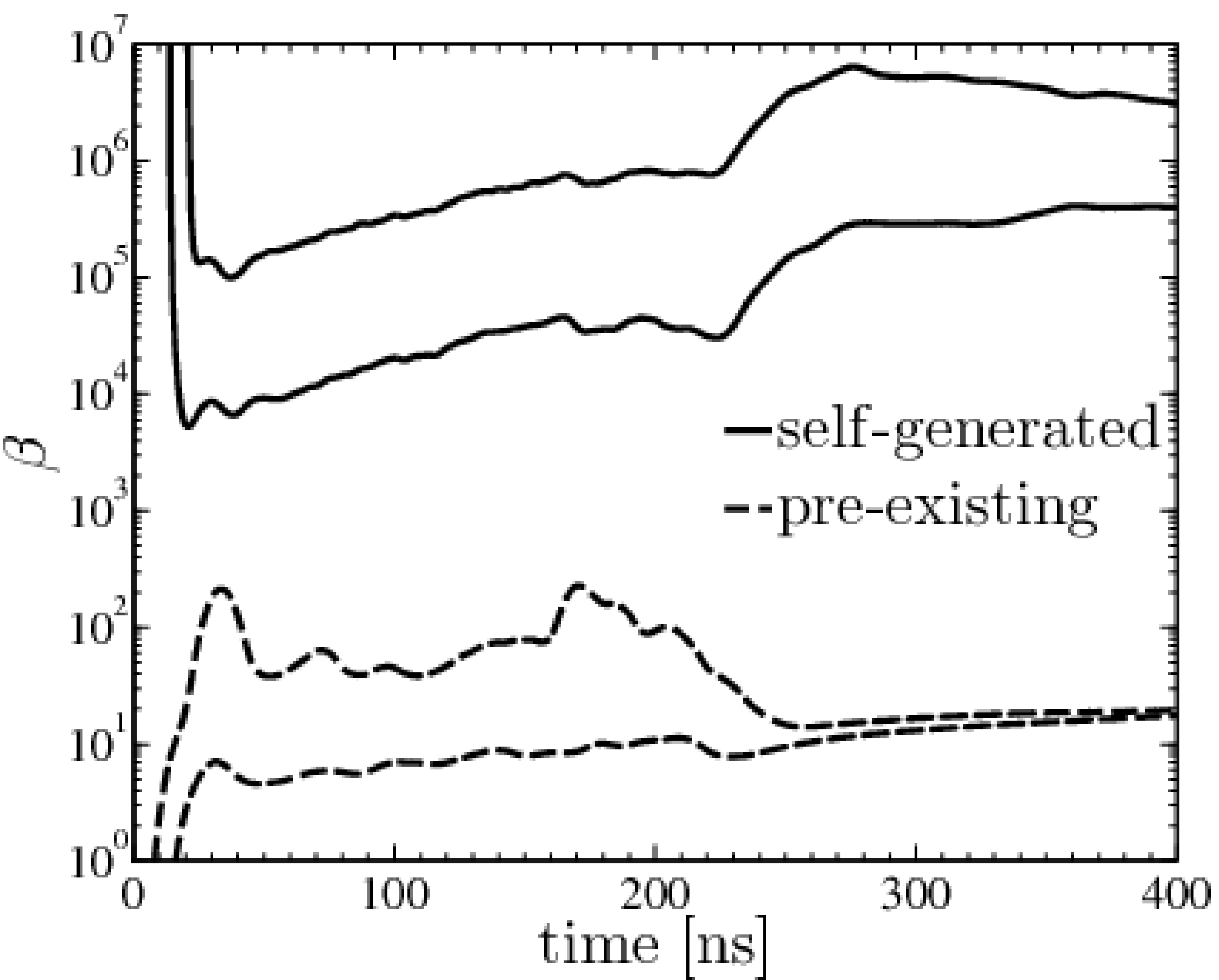}

	\caption{Magnetic field and plasma $\beta$ over time for the self-generating and pre-existing field cases. (top panel) Magnitude of the magnetic field. The thick curves indicate the arithmetic mean, while the thinner lines above and below show the 75th and 25th percentiles, respectively. For the pre-existing case, the initial mean magnetic field is amplified by a factor of 2 to 3 during the driving stage. The field then decreases when driving is halted. Note that the self-generating case produces only kilogauss fields. (bottom panel) Plasma $\beta$ values ($\beta=p/(8\pi B^2)$). The top and bottom lines for each case represent the 75th and 25th percentiles, respectively. The mean is not shown, as localized magnetic field voids produce small regions of extremely high $\beta$, skewing the results. The preexisting case generates $\beta$ values on the order of 1 to 100. The self-generating case generates very high $\beta$ values, resulting in a thermally dominated flow field.}
	\label{f:mag_statistics}
\end{figure}
While magnetic fields do not appear to play a role in the hydrodynamic development of the driven system, the role of generation and amplification in this driven turbulence scenario is interesting in its own right.

The evolution of the magnetic field strength and plasma $\beta$ for the two magnetized cases is shown in Fig.~\ref{f:mag_statistics}. For the case of a weak pre-existing magnetic field with $\beta_{ref}=20$, the initial magnitude of the magnetic field $5\times10^5$~G. During the driving phase the field is amplified by a factor of 2, with a mean field strength of $1\times10^6$~G. The distribution of field strength is roughly Gaussian during this period with a negative skew. The cessation of laser driving results in the distribution narrowing and the field strength decaying as the magnetic field is advected from the turbulent core.

The self-generating case quickly reaches its peak shortly after the TC is overrun by the first shocks and produces field strengths on the order of $1\times10^4$~G. The resulting field distribution is also a negative skew non-Gaussian. Unlike the pre-existing case, the shape of the distribution remains static post-driving. However, this distribution undergoes a shift as the magnetic field decays at the same rate regardless of field strength. 

The plasma $\beta$ distribution for the pre-existing case predominantly covers the moderate-to-weak field range of $10\le\beta\le 100$, which agrees with our initial magnetic field estimate of $\beta_{ref}=20$. This distribution contains fluctuations in the larger-$\beta$ region and remains roughly constant for smaller-$\beta$. In the post-driving phase $\beta$ decreases to a uniform value on the order of 10 throughout the domain. 

In contrast, the distribution for the self-generating case maintains the same structure throughout the driving phase, much like the distribution of $B$.  The bulk of the distribution is in the range of $10^4\le\beta\le 10^5$, indicating very weak effects on hydrodynamics due to the magnetic field. After driving $\beta$ increases drastically in contrast to the pre-existing field case. 

\begin{figure}[h!]
	\begin{center}
	
	\begin{overpic}[width=8cm]{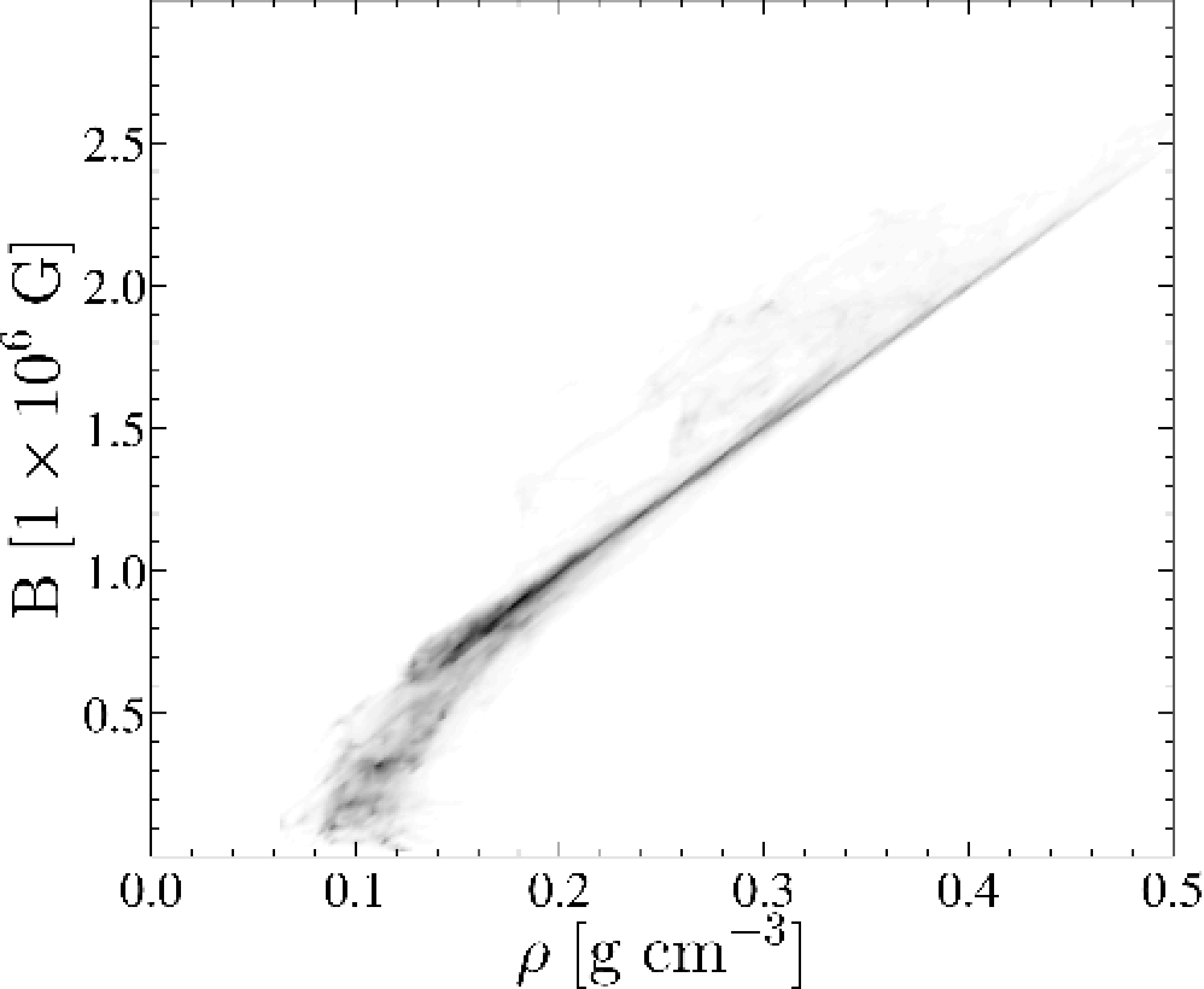}
		\put(20,75){(a) pre-existing field}
	\end{overpic}
	\begin{overpic}[width=8cm]{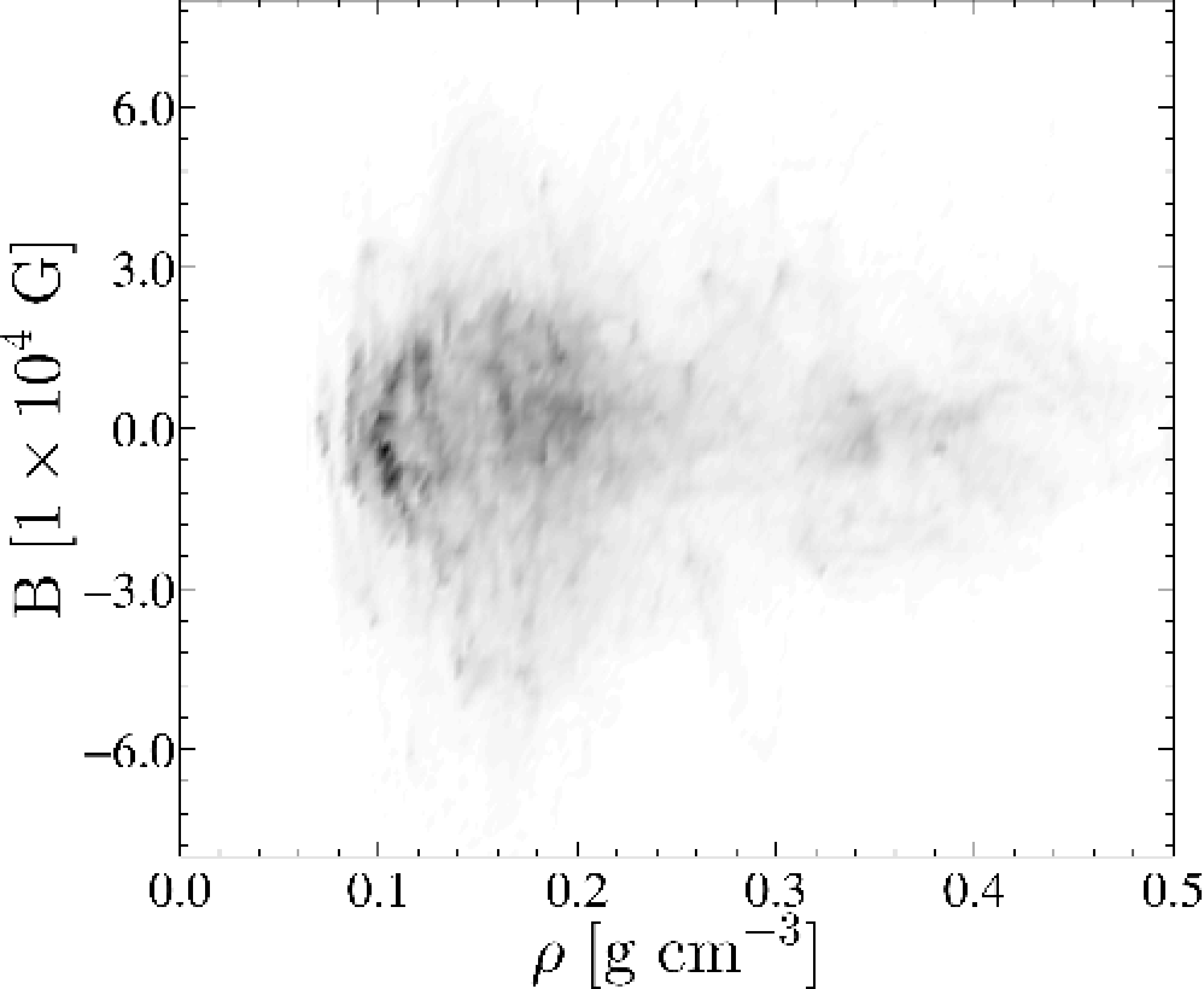}
		\put(22,75){(b) self-generated field}
	\end{overpic}
	  
	\end{center}
	\caption{Pseudocolor plot of the bivariate probability distribution of magnetic field strength and plasma density for the two magnetized cases. (a) The field strength for the pre-existing field correlates linearly with density, indicative of compression of the magnetic field lines as the gas compresses. (b) The self-generated field strength indicates that more complicated physics is involved than pure compression effects. There is no visible correlation between field strength and density for this case, indicating that amplification of field strength due to material compression is, at least, strongly suppressed.}
\label{f:magz_pdfs}
\end{figure}

Figure~\ref{f:magz_pdfs} shows the probability distributions of the magnetic field magnitude with respect to the density and gradient of density. For the pre-existing field case, there is a linear correlation between the magnetic field strength and the density. For compressible flows, this is indicative that the compression of the fluid is also compressing the field lines. In contrast, no correlation is seen between the density and field strength for the self-generating case. This does not indicate an absence of field line compression, but rather the complicated physics involved with generation hides such a correlation. Preliminary investigation into dependence for the self-generated case shows that short term, strong field events occur (on the order of megagauss), but it is difficult to quantify such behavior with the available data. We recommend further work in this area, with a focus on Lagrangian particle analysis.

\section{Hydrodynamic evolution of the self-generated case}
\label{s:selfgenerated}
\subsection{Morphology}
\label{s:morphology}
\begin{figure*}[ht!]
	\begin{center}

		\begin{overpic}[width=5.5cm]{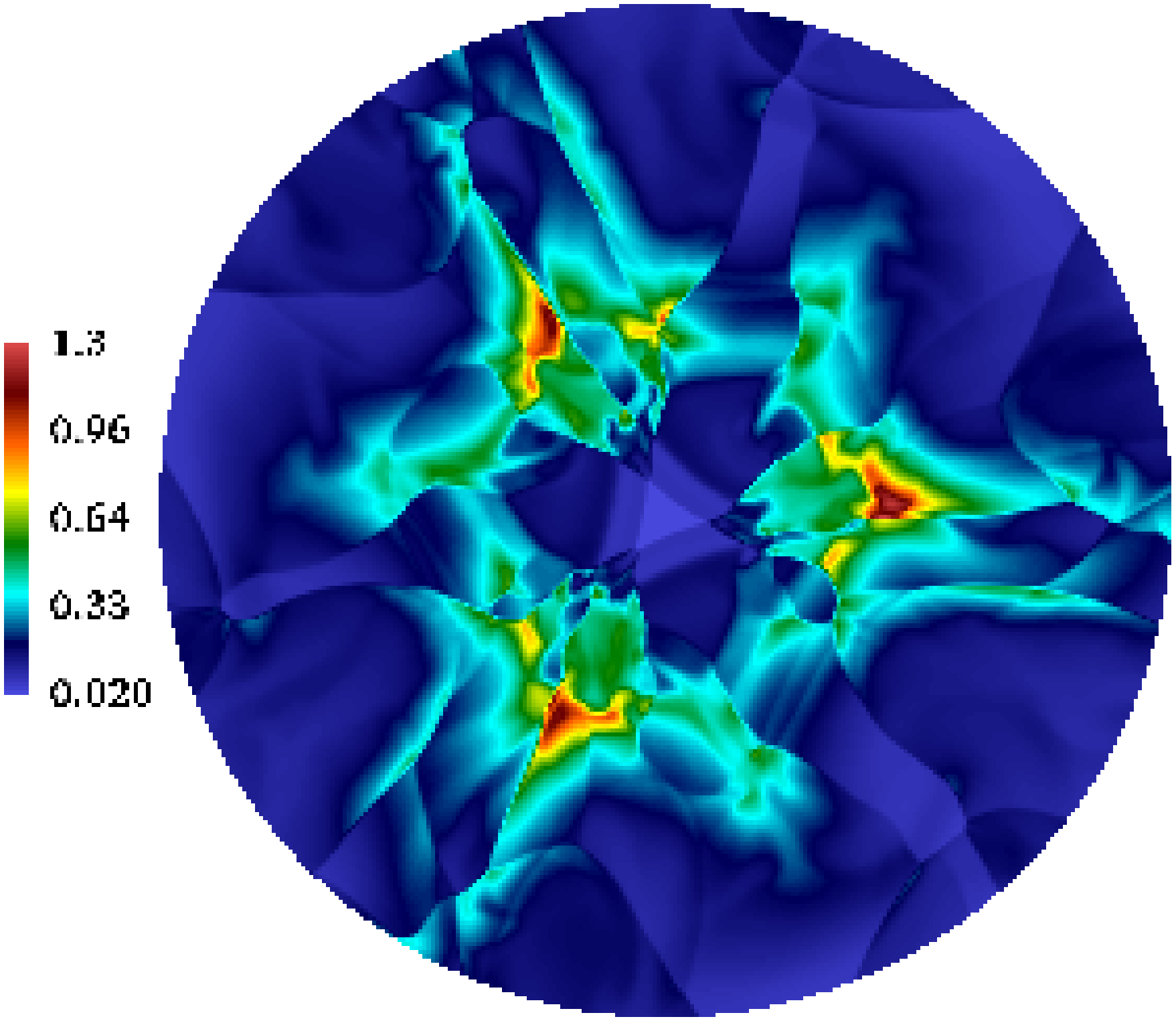}
			\put(40,-10){(a) $t=23$ ns}
		\end{overpic}
		~
		\begin{overpic}[width=5.5cm]{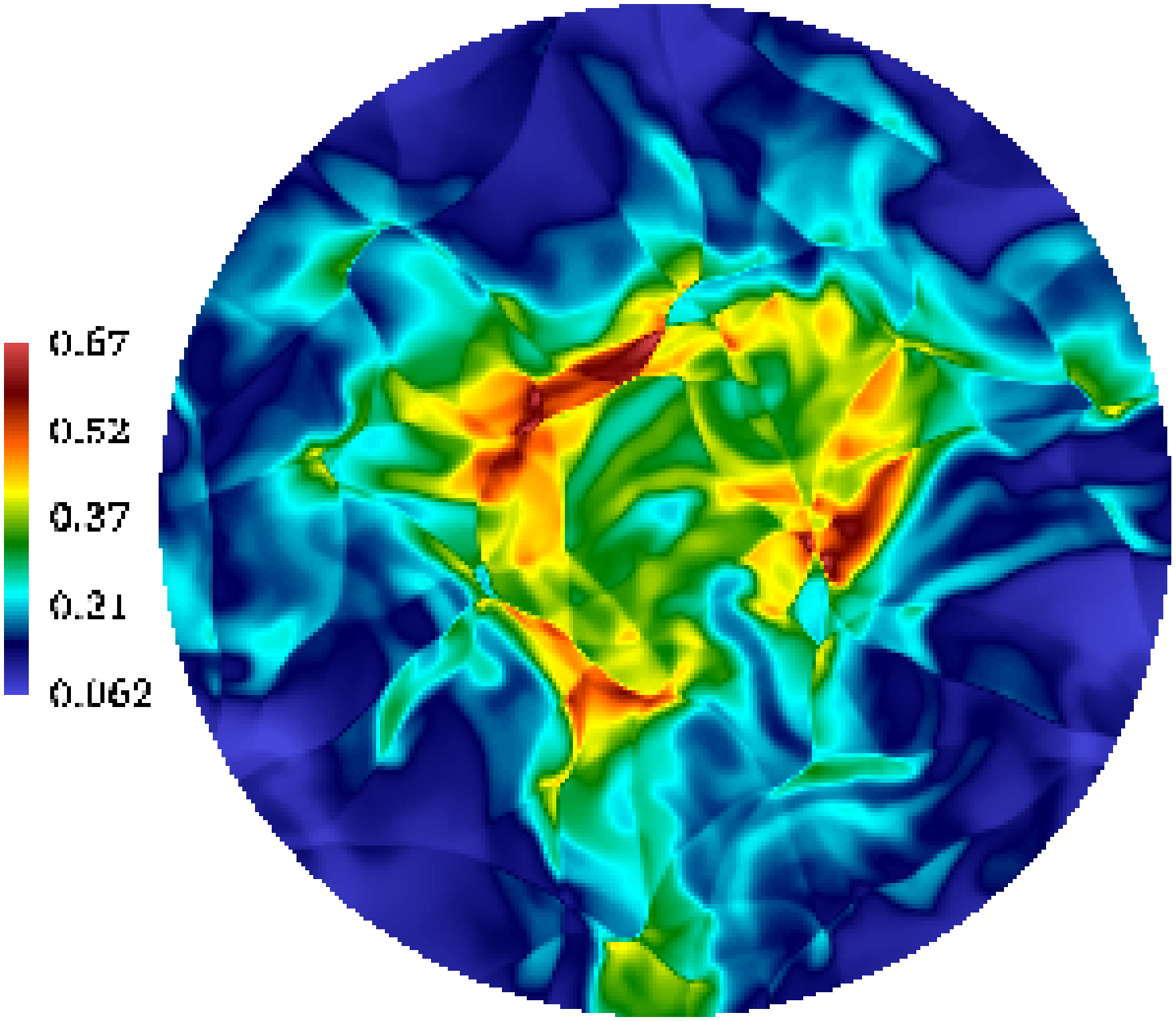}
			\put(37,-10){(b) $t=130$ ns}
		\end{overpic}
		~
		\begin{overpic}[width=5.5cm]{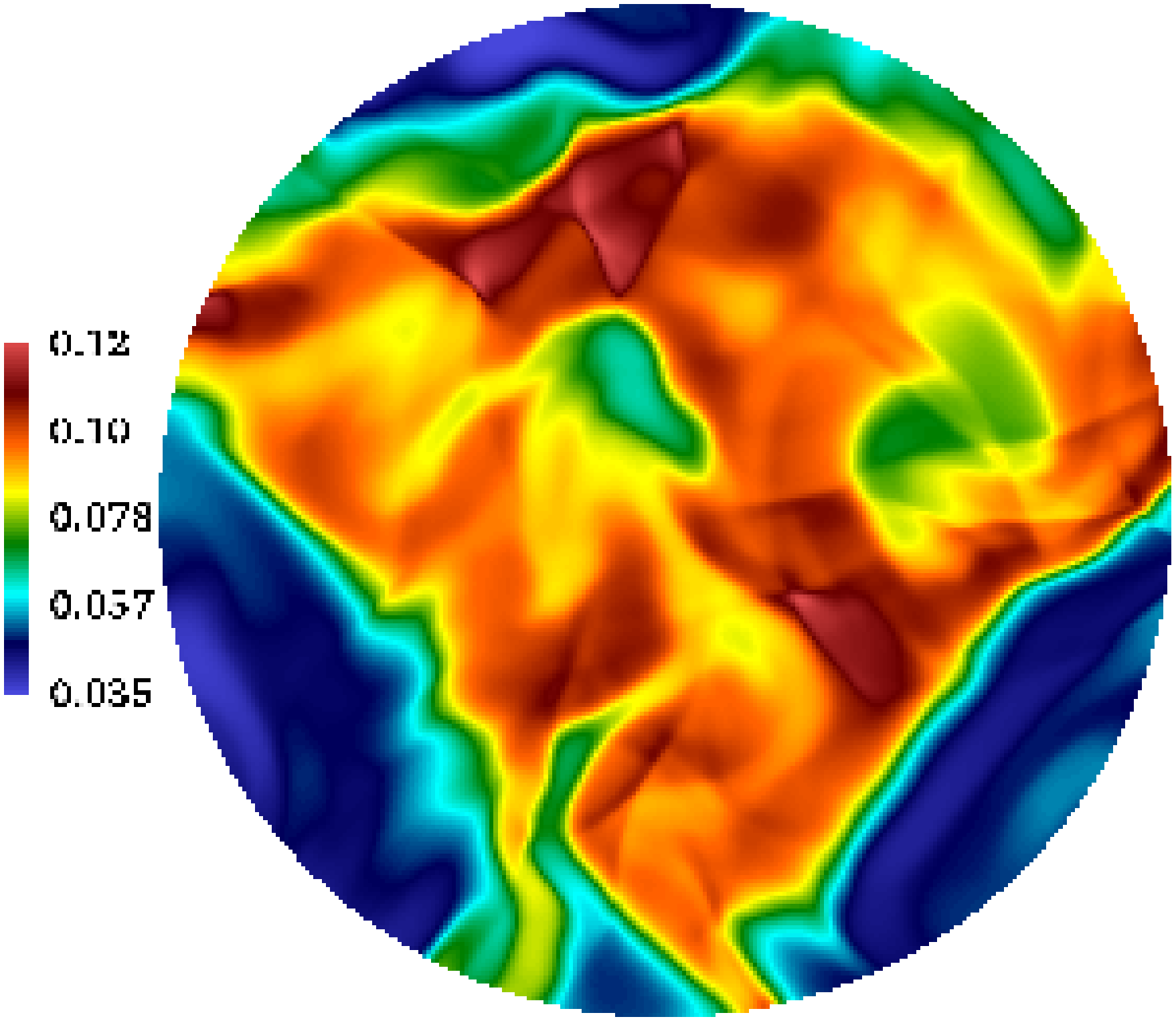}
			\put(37,-10){(c) $t=230$ ns}
		\end{overpic}
	\end{center}
\caption{Pseudocolor plots of density (g cm$^-{3}$) for the self-generated magnetic field case in the turbulent core. Mesh resolution in the region is 4\micron. (a) $t=23$~ns (b) $t=130$~ns (c) $t=230$~ns}
\label{f:2d-self_density}
\end{figure*}
Pseudocolor plots of the density in the region of interest at multiple times is depicted in Fig.~\ref{f:2d-self_density}. At $t=23$~ns the initial laser-driven shocks have penetrated the light core material. The small triangular region at the center of the TC is still unshocked and the preheated region in front of the hydrodynamic shocks can be seen. There are no discernible fluid instabilities at this time, although at previous times small Rayleigh-Taylor and Kelvin-Helmholtz instabilities could be seen before being overrun by later shocks. The interaction of initial shocks has compressed the triple point areas by a factor of 13 from the shell density. 

In the middle of the driving phase ($t=130$~ns) confinement due to the laser arrangement results in a dense core embedded in a lower density medium. The flow structure appears very chaotic, with the multitude of shocks passing through the core clearly visible. This image is representative of the remaining 70~ns of driving. 

At the peak of the post-driving phase (230~ns), the density field exhibits interesting flow features. In particular, there appears to be low density bubbles embedded in the more dense remnant of the core. We note that the maximum density has decreased by a factor of nearly 6 from the snapshot at $t=130$~ns. The bulk distribution of density appears (visually) to be quite anisotropic, with a low density imprint at 120\degree, 220\degree, and 320\degree. Note that the last laser triplet was fired nearly $t=30$~ns prior, and the resulting shocks have left the domain already.

\subsubsection{Isotropy}
\begin{figure*}[ht!]
	\begin{center}
	
	
	\begin{overpic}[width=5.5cm]{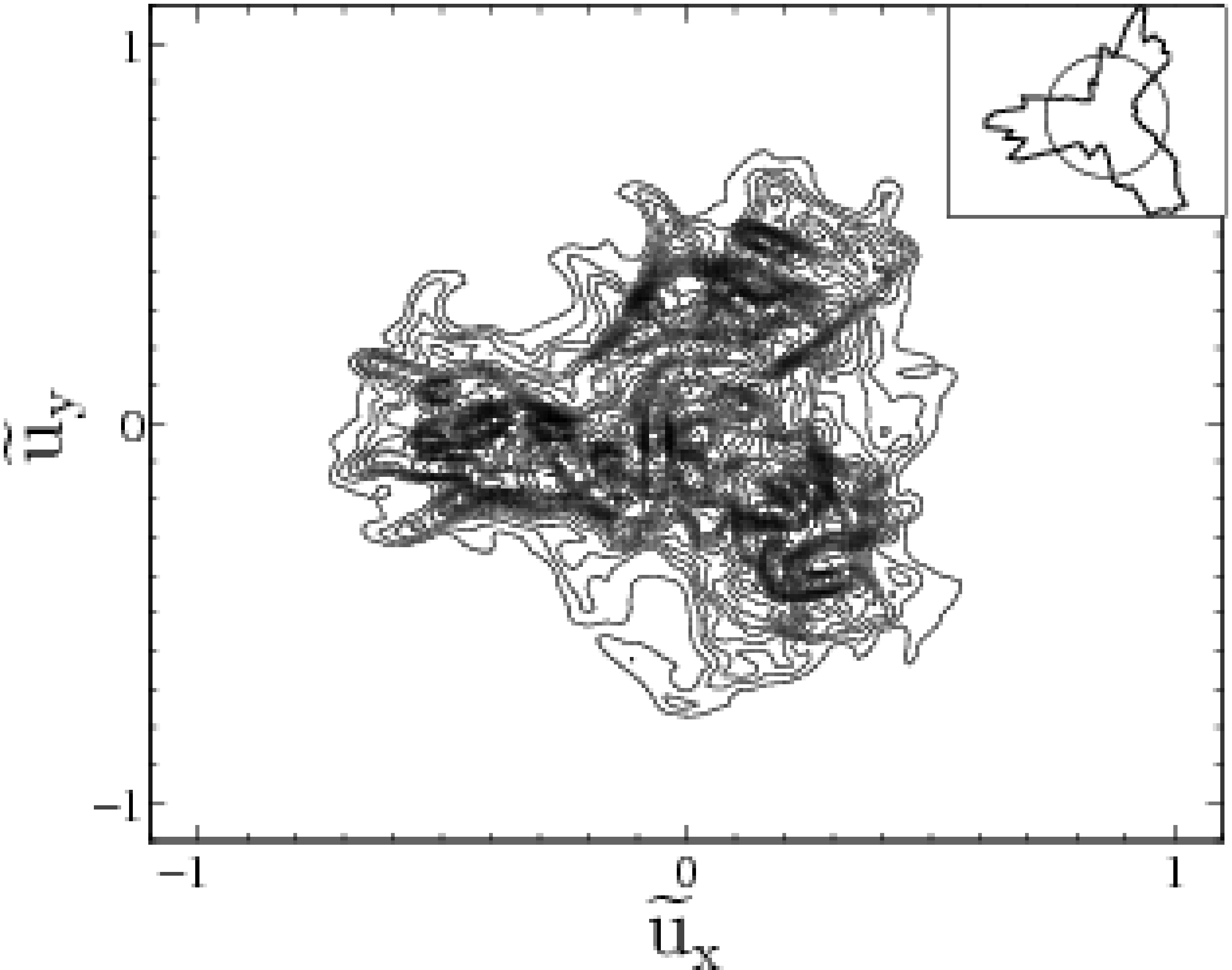}
		\put(15,13){(a) $t=30$ ns}
	\end{overpic}
	~
	\begin{overpic}[width=5.5cm]{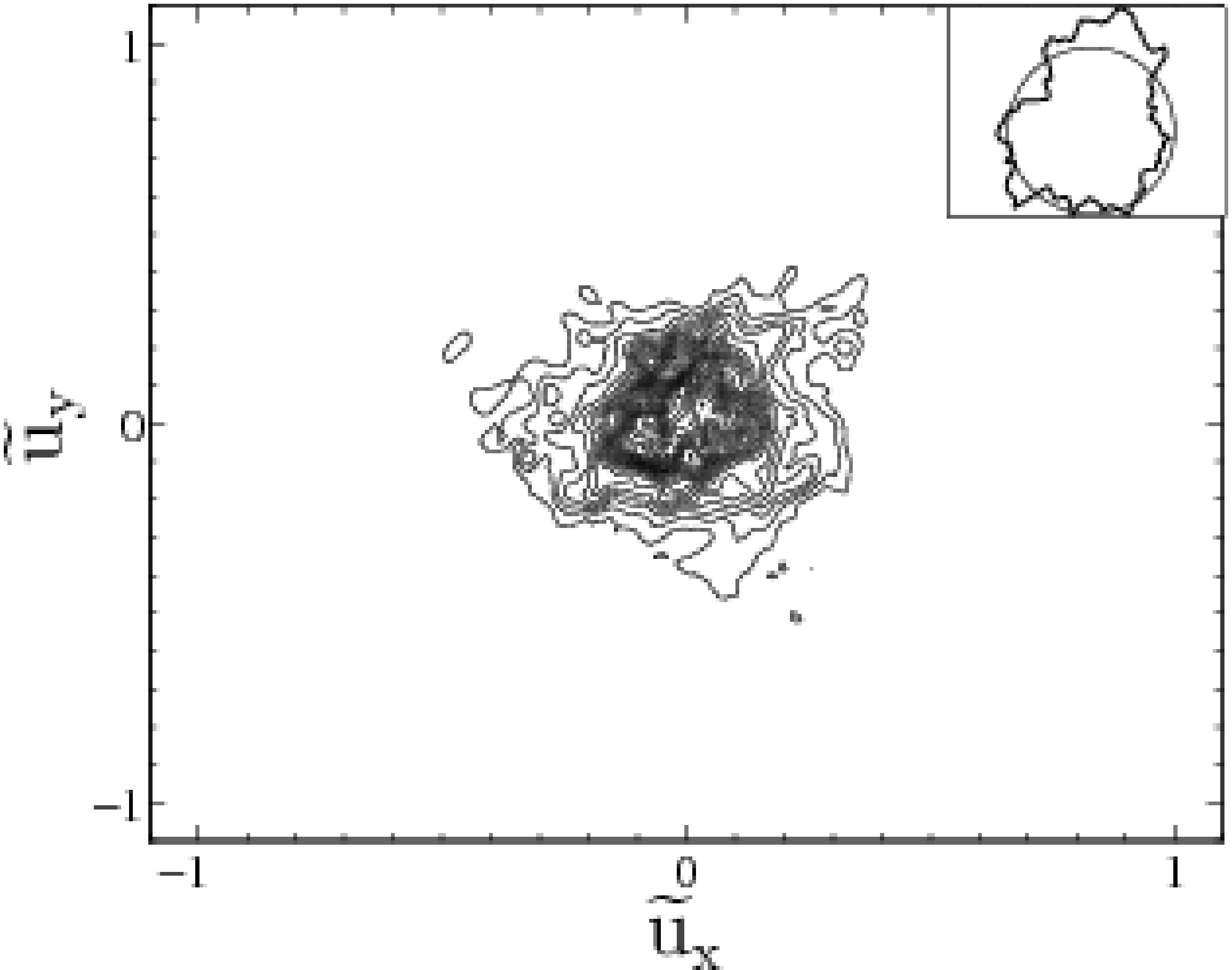}
		\put(15,13){(b) $t=130$ ns}
	\end{overpic}
	~
	\begin{overpic}[width=5.5cm]{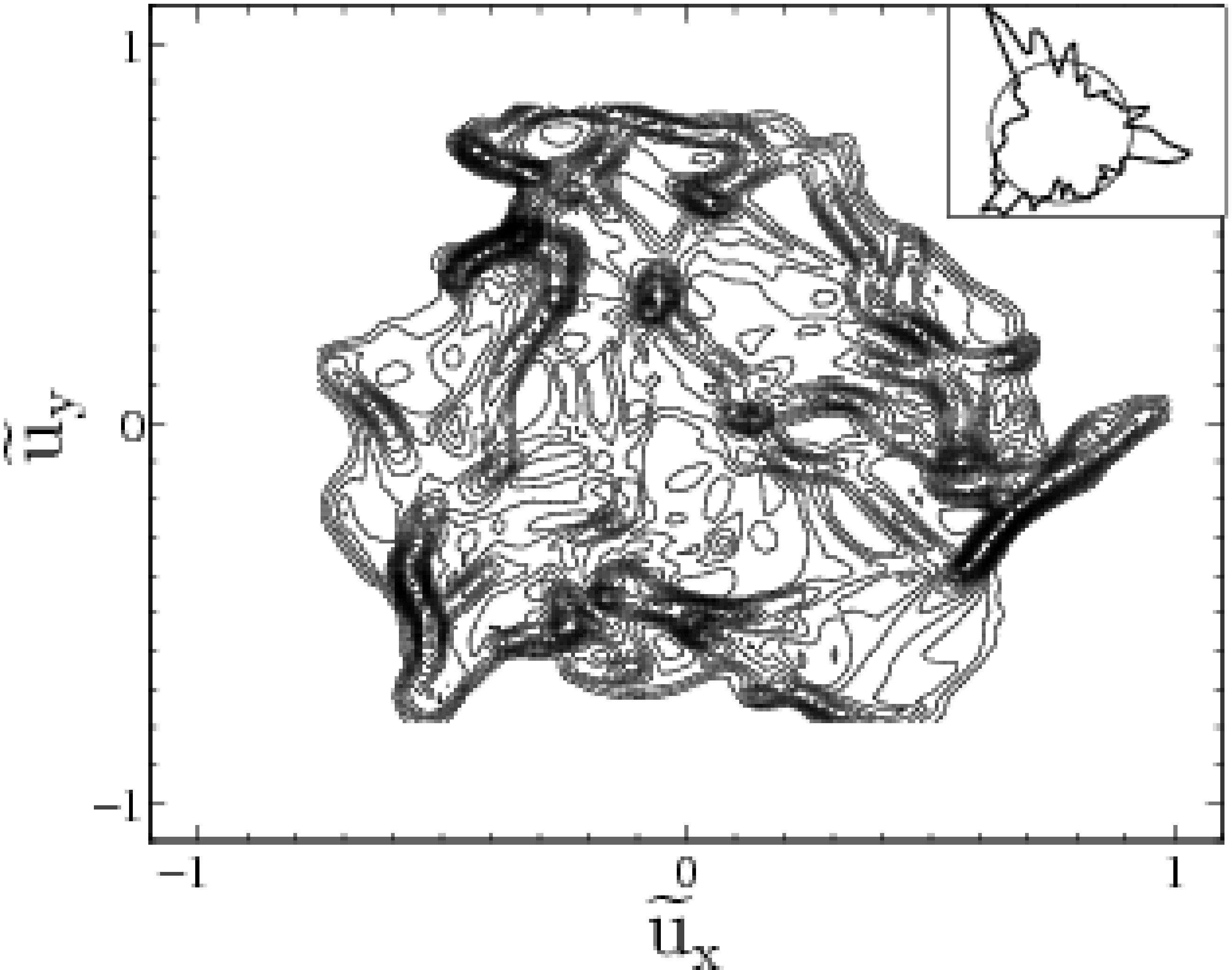}
		\put(15,13){(c) $t=230$ ns}
	\end{overpic}
	
	\end{center}
	\caption{
	Isocontours of the bivariate probability distribution for scaled velocity vector components in the turbulent core. The individual components of velocity for each grid cell are scaled by the maximum velocity magnitude at each snapshot to generate $\widetilde{u}_{x/y}$. The insets show a polar histogram indicating directionality of the velocity field. The solid circle shows how a uniform (isotropic) distribution of velocity would appear. (a) $t=30$~ns, $|{\bf u}|_\mathrm{max}=9.3\times10^{6}$ cm~s$^{-1}$. The initial imprint from the first laser triple is clearly visible. (b) $t=130$~ns, $|{\bf u}|_\mathrm{max}=9.4\times10^6$ cm~s$^{-1}$. During the driving phase, the velocity field achieves an isotropic distribution. (c) $t=230$~ns, $|{\bf u}|_\mathrm{max}=5.5\times10^6$ cm~s$^{-1}$. The absence of confining ram pressure allows the compressed plasma to expand when laser driving ceases, which manifests as the of the velocity space.
	}
	\label{f:2d-self_isotropy}
\end{figure*}
The amount of physical material available in the experimental setting is limited due to the balance between target size and available driving energy. Consequently, confinement of the shocked plasma in the region of interest is of great benefit to the experiment. Therefore, we investigate the isotropy of the velocity field over time, with the most beneficial outcome being that the velocity evolves to a fully isotropic state where the bulk momentum is zero. Additionally, isotropy of the flow field has additional implications for turbulence and the application of Kolmogorov theory.

Probability density contours of the velocity components in the region of interest are shown in Fig.~\ref{f:2d-self_isotropy}. The axes are scaled by the maximum instantaneous velocity magnitude to standardize the figures. The inset shows the angular probability distribution mapped to polar coordinates with the solid line. A fully isotropic distribution is shown with the dashed circle. 

As flow features (such as shocks) penetrate the TC and evolve, large velocity gradients are formed. This behavior is reflected in velocity space by tightly packed isocontours. Additionally, the velocity \emph{within} fluid structures has relatively mild gradients, leading to high probability regions. Therefore, the distribution of isocontour lines is a direct result of the fluid structures inside of the region and one can infer information about the aggregate behavior of flow features in the domain.

At $t=30$~ns three ``ejecta'' can be seen extending from the origin. In physical space this corresponds to the three dense, multiply shocked regions in Fig.~\ref{f:2d-self_density}. This distribution is highly anisotropic, which is illustrated in the inset. Note that the inset probability plot is rotated by 180\degree~from the apparent position of the shocks because it is a representation of the direction of motion, not position.

In the middle of the driving phase ($t=130$~ns), the core has been overrun repeatedly with shocks and the chaotic motion induced by the laser driving results in a much more isotropic distribution of velocities shown in the inset. The circular banding of isocontours around the origin implies that flow structures are causing a roughly isotropic distribution of velocity gradients. This indicates that the shocks (in particular, their normals) are isotropically distributed throughout the domain, as the primary source of acceleration in this system comes from the kinetic, laser-driven material (since magnetic field effects play a minor role).

The final snapshot at $t=230$~ns illustrates that the velocity components have begun to fill out the velocity space. There are strong, localized flow features that result in directionally biased velocities; however, the bulk of the distribution remains isotropic. The localized features are the result of the final shocks passing through the domain. The bulk outward motion is due to the lack of laser-driven confinement which allows for pressure gradients to drive material away from the origin.

\subsection{Flow field structure during driving}

\begin{figure*}[ht!]
	\begin{center}


		\begin{overpic}[bb=0   0 1060 1060, clip, height=6.5cm]{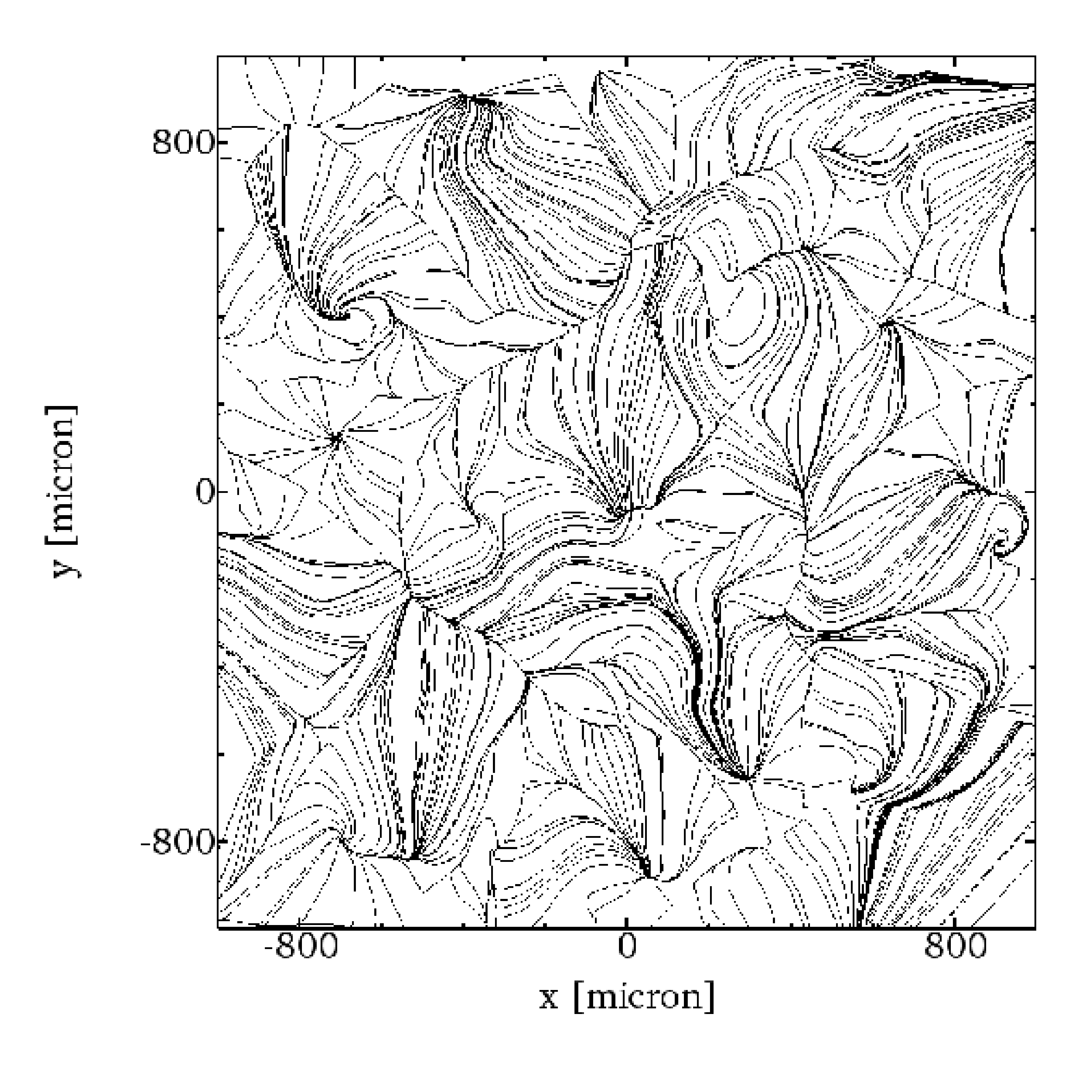}
			\put(21,90){(a)}
		\end{overpic}
		~
		\begin{overpic}[bb=128 0 1060 1060, clip, height=6.5cm]{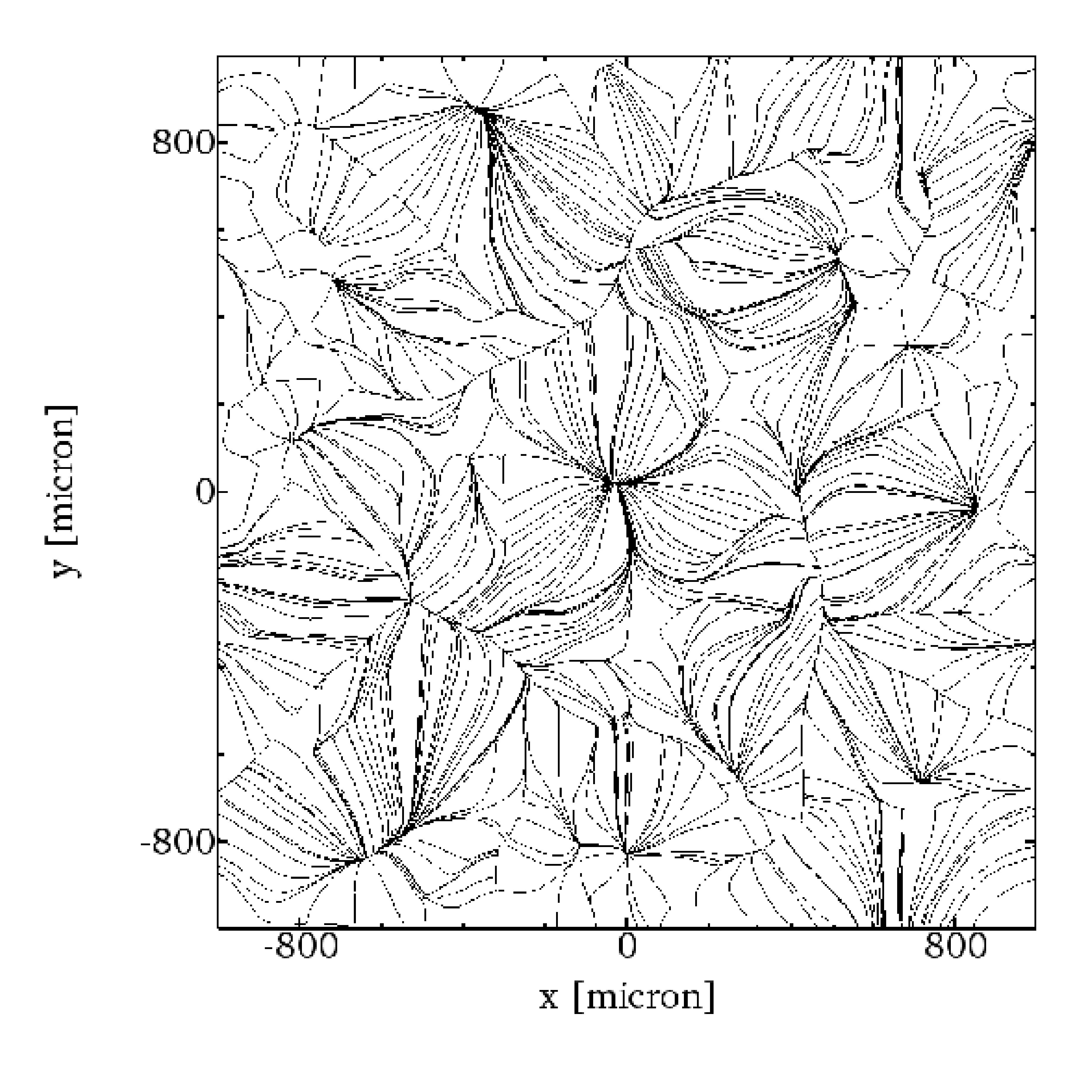}
			\put(10,90){(b)}
		\end{overpic}
		~
		\begin{overpic}[bb=128 0 1060 1060, clip, height=6.5cm]{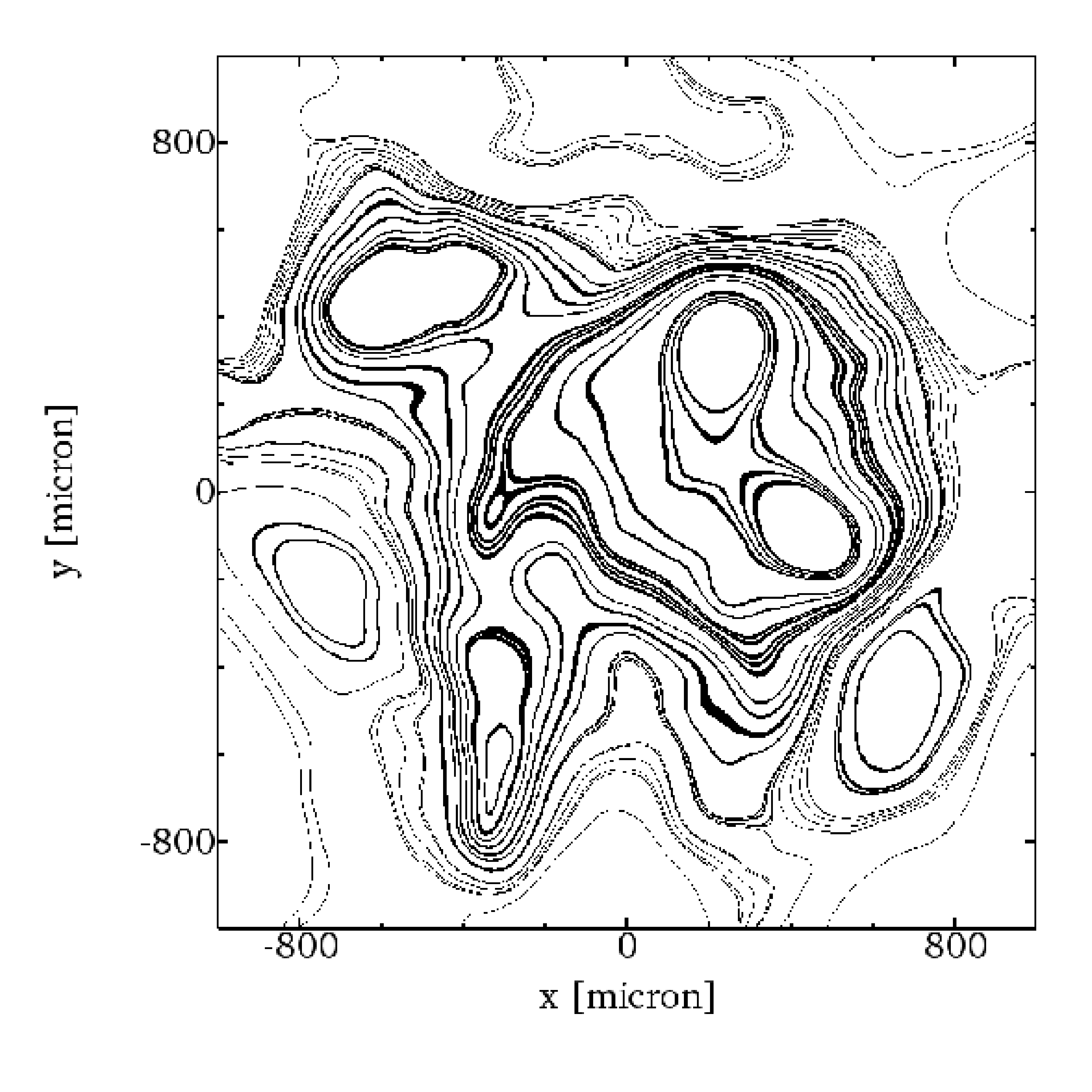}
			\put(10,90){(c)}
		\end{overpic}
		
		\end{center}
	\caption{Streamlines of the velocity field and its decomposition at $t=130$~ns. This square region is centered in the TC and is the area which the spectra is computed on. (a) Total velocity. Note the combination of point sources, link sinks, and vortical motions. (b) Compressive velocity component. There are multiple point sources visible, indicating expanding regions. The coalescing streamlines in the domain are line sinks, and correspond to compression of the material. (c) Solenoidal velocity component. The underlying vortical flow structure is showcased, with a nested structure to the vortices. Note that for two dimensional turbulence, the smaller vortices are likely inducing the formation of the larger vortical cells.}
	\label{f:streamlines}
\end{figure*}
Additional information can be obtained about the flow field during the driving phase by decomposing the velocity into constituent components. Using the Helmholtz theorem \cite{arfken+05}, we can reformulate the velocity field as $\bf{u} = \bf{u}_c + \bf{u}_s + \bf{u}_h$, where $\bf{u}_c$, $\bf{u}_s$, and $\bf{u}_h$ are the compressive, solenoidal, and harmonic components, respectively. We do not consider the harmonic component in our analysis as it exists only as a correction for non-periodic domains and plays no dynamical role in the system. In order to compute these terms, we rewrite them in terms of their Fourier transforms,
\begin{equation}
\label{e:ucFourier}
\bf{\hat{u}}_c\left(\bf{k}\right)=\left[\bf{k}\cdot\bf{\hat{u}}\left(\bf{k}\right)\right]\bf{k},
\end{equation} 
\begin{equation}
\label{e:usFourier}
\bf{\hat{u}}_s\left(\bf{k}\right)=\left[\bf{k}\times\bf{\hat{u}}\left(\bf{k}\right)\right]\times\bf{k}. 
\end{equation}
Equations \ref{e:ucFourier} and \ref{e:usFourier} are easily computed and inverted to produce the corresponding velocity fields in physical space. Additional details of our computations involving the Fourier transform are found in Sec.~\ref{s:spectra}.

Figure~\ref{f:streamlines} shows streamlines of the velocity field and its components at $t=130$~ns for a square subsection of the region of interest. The left most image shows the streamlines for the total velocity field ($\bf{u}$).

The middle image illustrates the compressive component of the velocity field ($\bf{u}_c$). The sources and sinks in the field are now clearly visible. The point-like areas where streamlines fan out are indicative of the expansion of material ($\nabla\cdot\bf{u}>0$). 

The domain is also populated by ``line sinks'' where streamlines abruptly end along a curve. These features are regions of fluid compression ($\nabla\cdot\bf{u}<0$), induced by either strong perturbations or weak shocks. Strong shocks in the domain are rare, and the flow field is predominantly populated by strong perturbations. 

The final set of streamlines are for the solenoidal component of the velocity field ($\bf{u}_s$). One can easily identify the underlying vortical structure of the flow field now. In particular, the domain consists of nested vortices typical of turbulence. As we are dealing with two dimensional turbulence, it is conceivable that the small vortices induce the larger structures due to the inverse enstrophy cascade. For the subsection pictured, the largest cell is approximately 1600\micron in diameter, while the smallest cell contained within has a diameter of roughly 100\micron. 

\subsubsection{Radial distribution}
\begin{figure}[h!]
	\begin{center}
		
		
		\begin{overpic}[width=4cm]{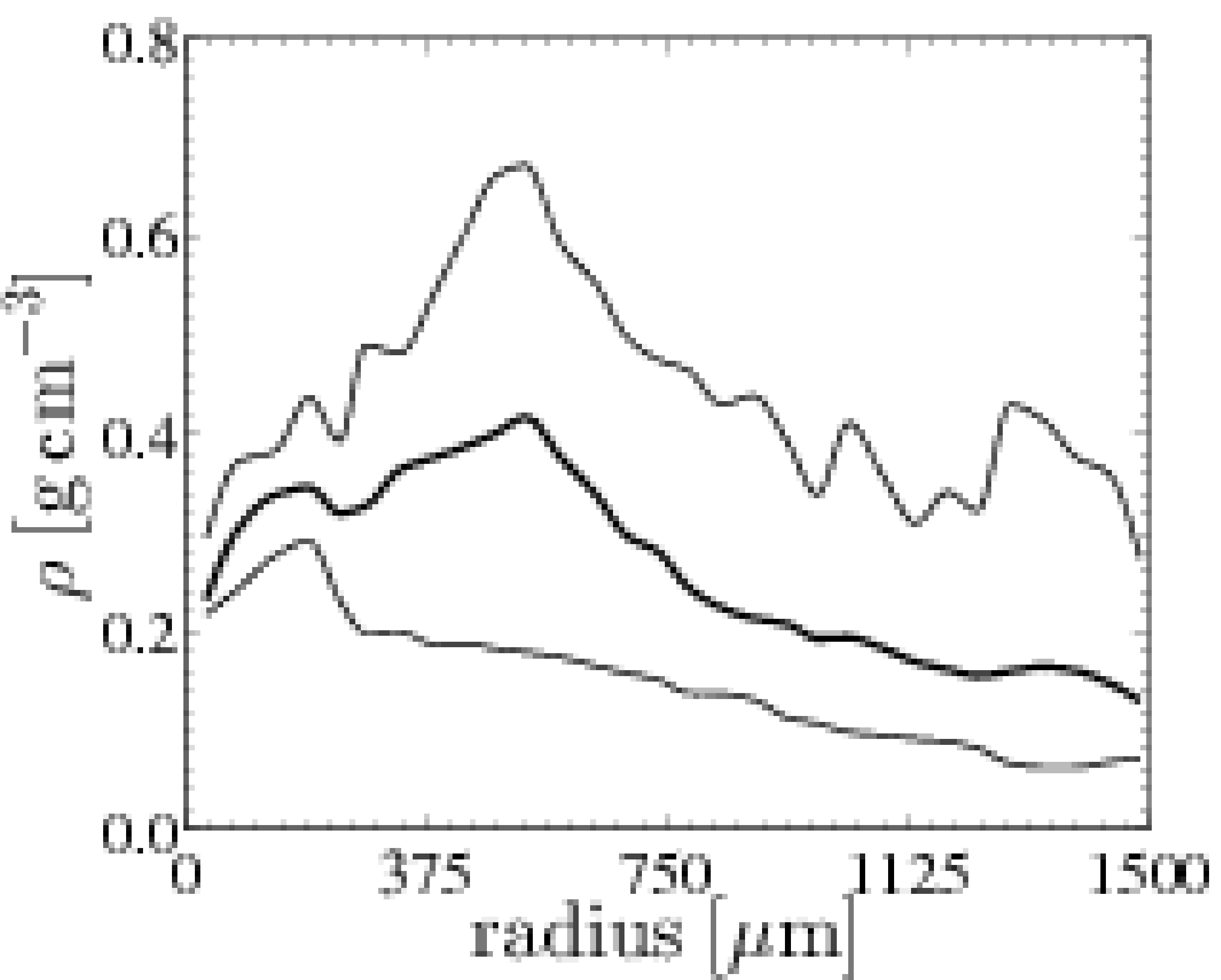}
			\put(55,67){$t=130$ ns}
		\end{overpic}
		~
		\begin{overpic}[width=4cm]{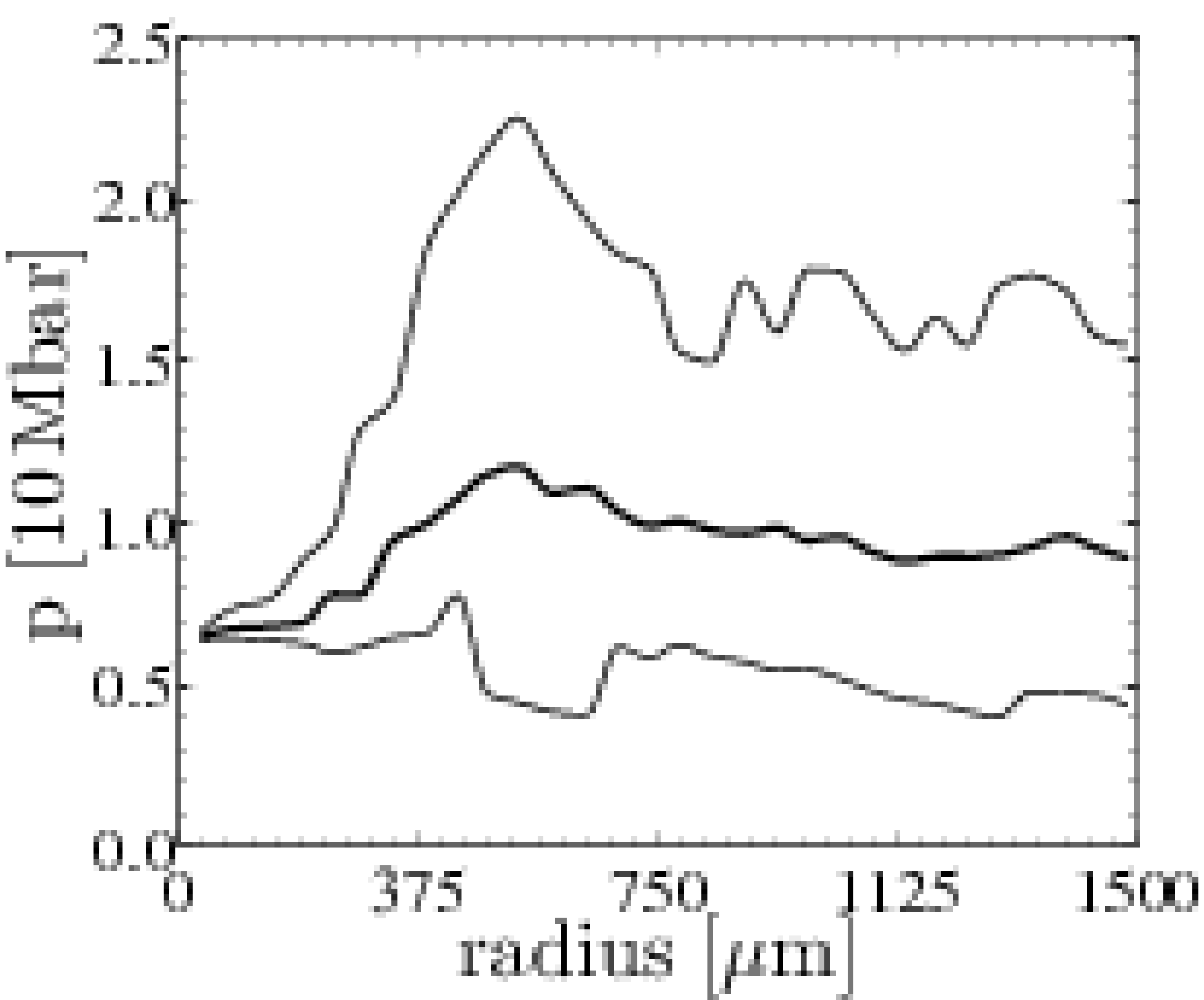}
			\put(55,67){$t=130$ ns}
		\end{overpic}
		\\
		\begin{overpic}[width=4cm]{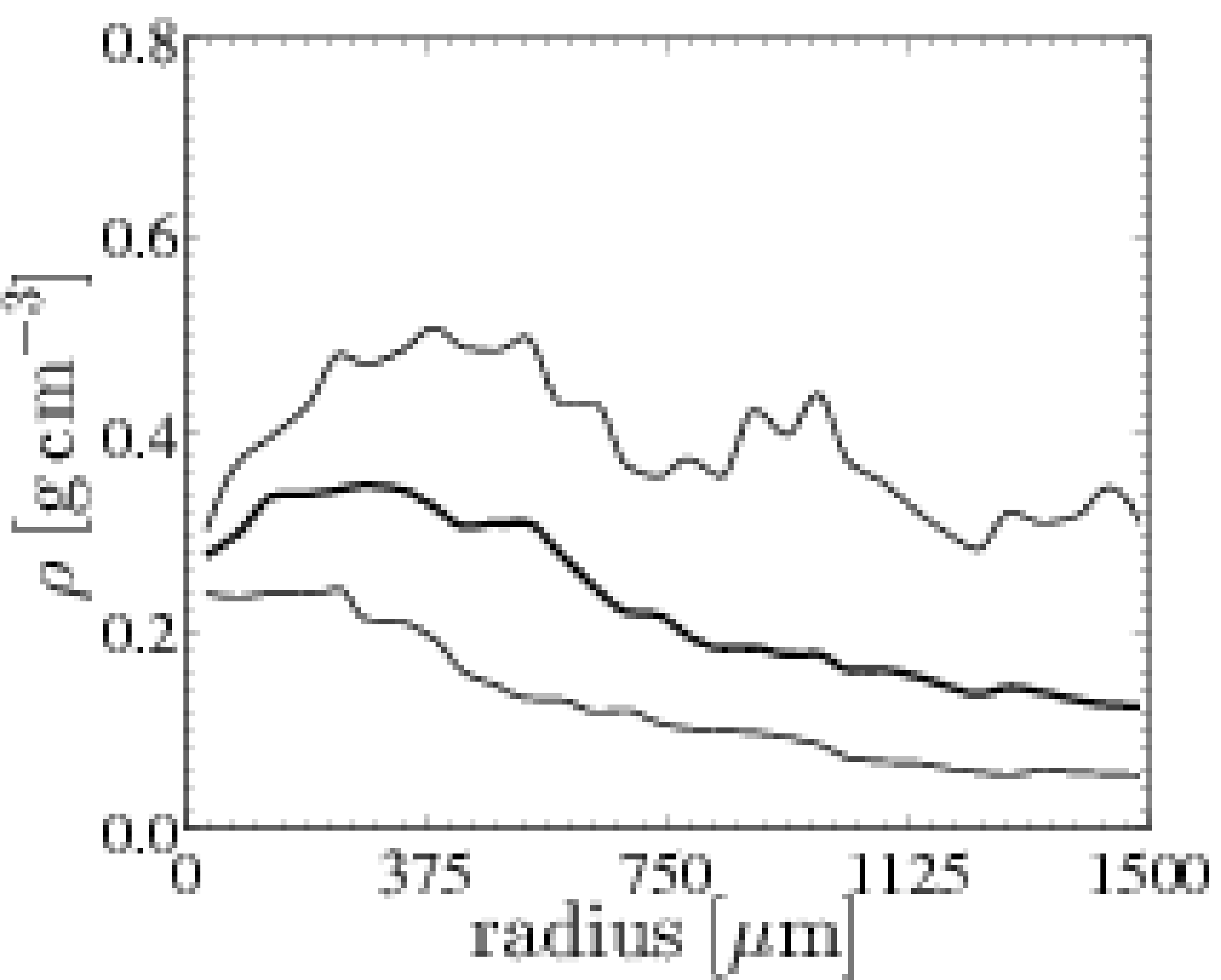}
			\put(20,67){$t=200$ ns}
		\end{overpic}
		~
		\begin{overpic}[width=4cm]{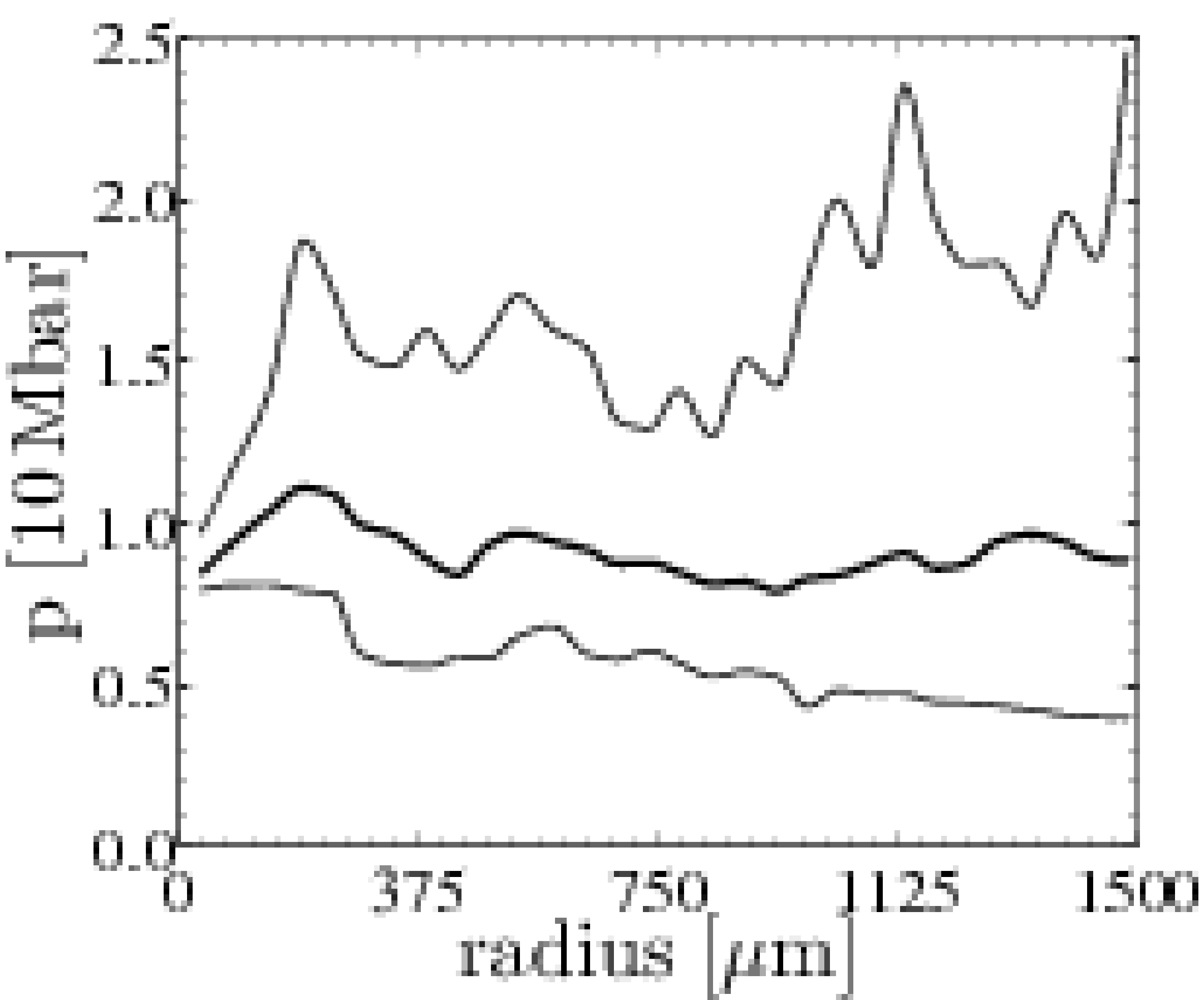}
			\put(20,67){$t=200$ ns}
		\end{overpic}
		\\
		\begin{overpic}[width=4cm]{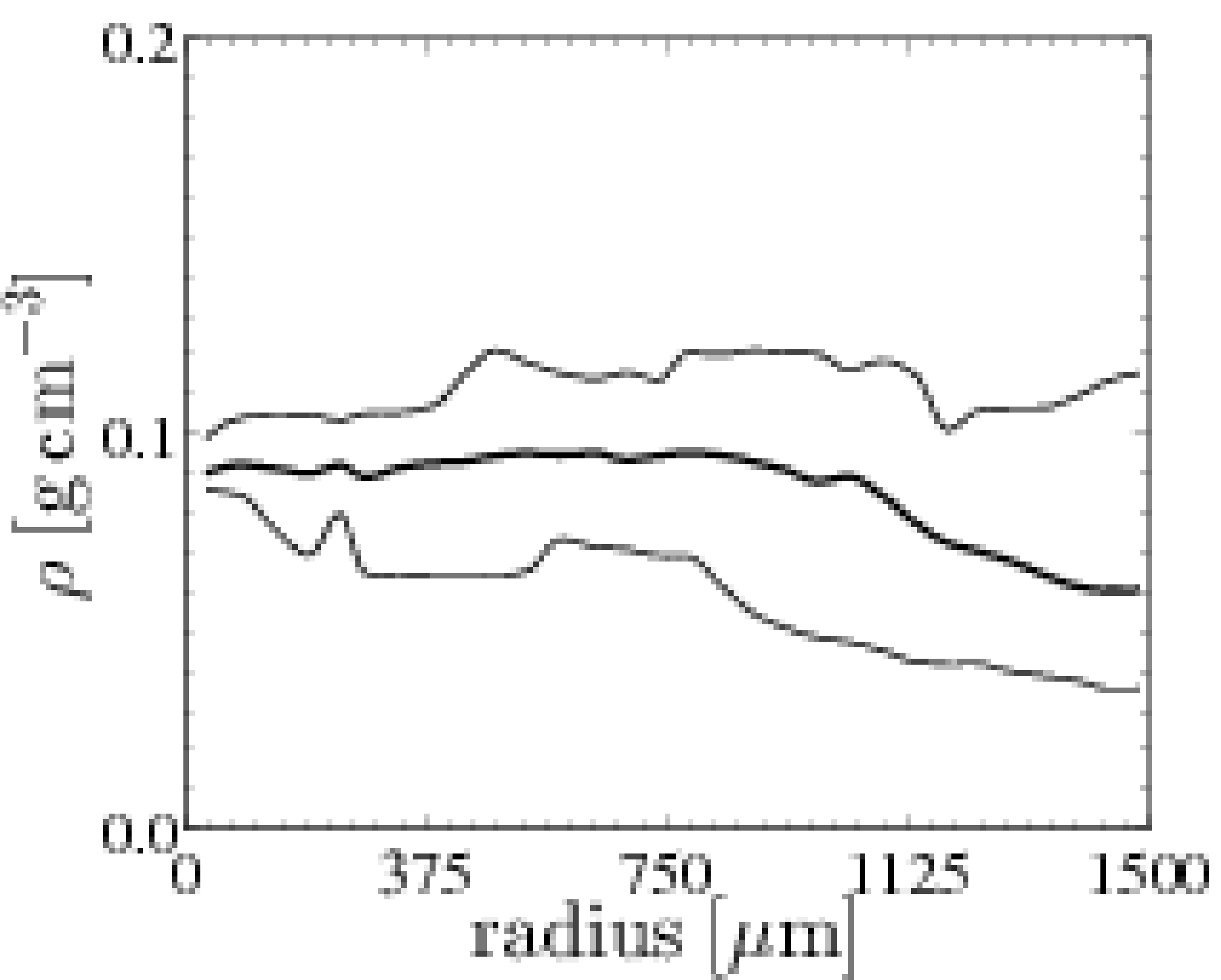}
			\put(20,67){$t=230$ ns}
		\end{overpic}
		~
		\begin{overpic}[width=4cm]{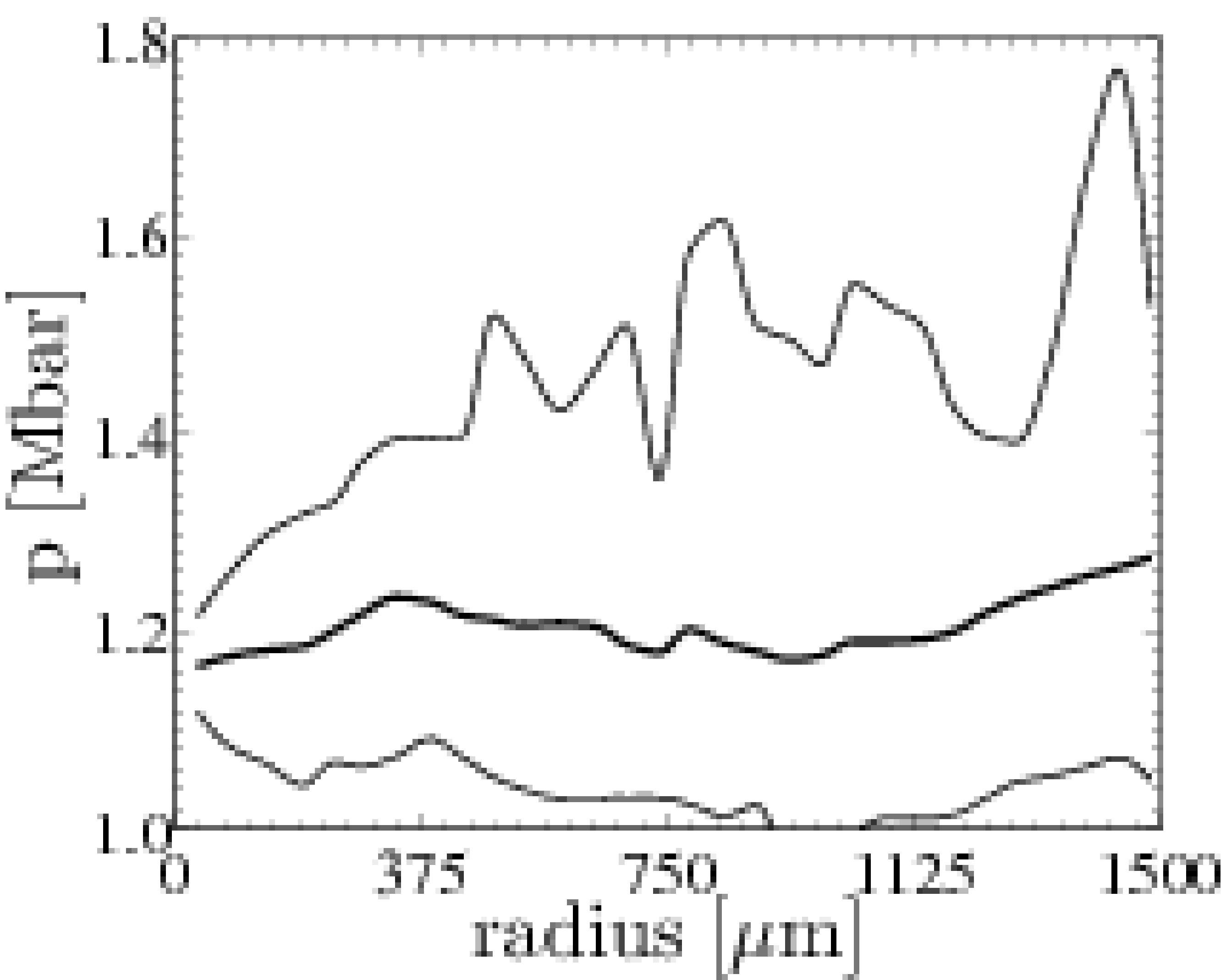}
			\put(20,67){$t=230$ ns}
		\end{overpic}

	\end{center}
	\caption{Radially averaged distributions of density (left column) and pressure (right column) inside the turbulent core at $t=130$~ns, $t=200$~ns, and $t=230$~ns. Thick solid lines denote the mean value while thin solid lines show the minimum and maximum values.}
	\label{f:2d-self_radialdists}
\end{figure}
Figure~\ref{f:2d-self_radialdists} quantifies the descriptions provided in Sect.~\ref{s:morphology}, showing density and pressure as a function of distance from the center of the target. These results are obtained by constructing probability distribution functions for grid cells contained in circular shells of 50\micron width. Thick solid lines denote the mean value while thin solid lines show the minimum and maximum values.

During the middle of the driving phase ($t=130$~ns), the density peaks at $600$\micron away from the core, and decays as the
radius increases. The pressure is relatively low at the center of the core, and increases until $600$\micron when it peaks and decays slightly. 

By the time the laser drives cease firing at $t=200$~ns, the density profile has flattened near the core, but continues to decay with radius. The pressure near the center of the core has increased, bringing the region of interest into approximate thermal equilibrium. 

At $t=230$~ns, at the peak of the rms-Mach number, the average density profile in the core has homogenized, staying roughly constant out to $800$\micron. Note that the maximum mean density has decreased by approximately a factor of 4. While the pressure profile indicates that the region is still roughly in thermal equilibrium, the mean value has decreased by an order of magnitude.

\subsubsection{Density probability distribution functions}
\begin{figure}[h!]
	\begin{center}
		\includegraphics[width=8cm]{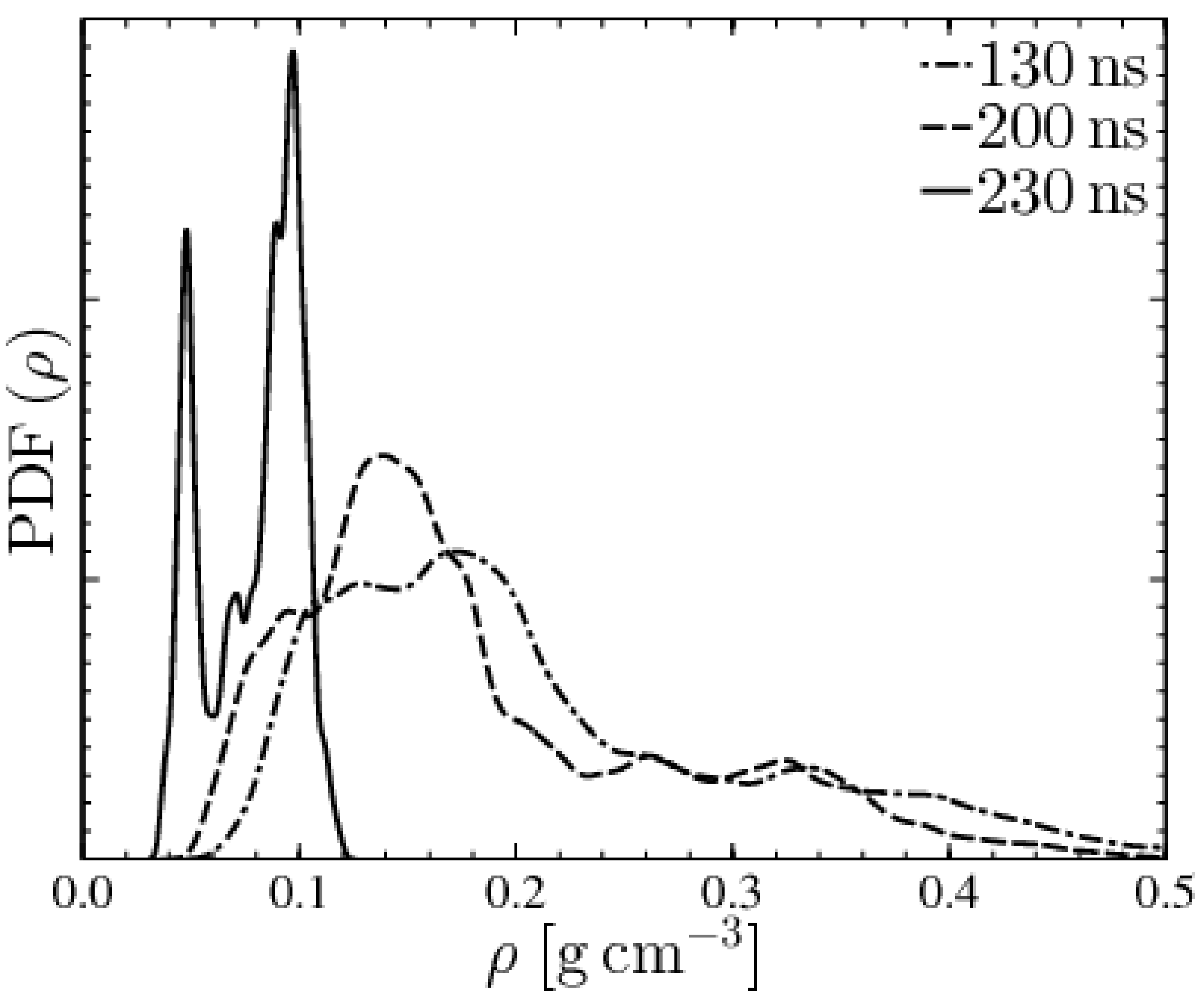}
	\end{center}
	\caption{Probability distribution function for the density in the turbulent core at $t=130$~ns, $t=200$~ns, and $t=230$~ns. The distribution at $t=230$~ns is scaled by 1/2 for visualization purposes. During the driving phase the density follows a lognormal distribution. However, once laser driving ceases the distribution becomes bimodal. 
	}
	\label{f:2d-self_pdfs}
\end{figure}
In addition to investigating the spatial dependence on density, we can also look at the probability distributions in the region of interest. Figure~\ref{f:2d-self_pdfs} shows these distributions during and after the driving phase for the self-generating case. Comparing the distributions at early and late times during driving, we note that the profiles are generally similar and follow log-normal distributions with parameters $\mu_{130} = -1.6435$, $\sigma_{130} = 0.4571$ for the $t=130$~ns distribution, and $\mu_{200} = -1.8020$, $\sigma_{200} = 0.4619$ for $t=200$~ns. These long tails are the byproduct of shock-driven turbulence. The late-time distribution also indicates that mass has been added to the region from the driving, but this was to be expected due to the nature of the open system. 

The post-driving distribution of density is quite different from that of the driving phase. It has become strongly bimodal and no longer exhibits the extended tail that characterized the driving phase distributions. In addition, the densities have shifted to lower values with the low and high density peaks at approximately $4.5\times10^{-2}$ g~cm$^{-3}$ and $9.4\times10^{-2}$ g~cm$^{-3}$, respectively. This result quantifies the structures at late times shown in Fig.~\ref{f:2d-self_density}.

\subsubsection{Kinetic energy power spectra}
\label{s:spectra}

\begin{figure*}[htbp!]
	\begin{center}
		

		\begin{overpic}[width=5cm]{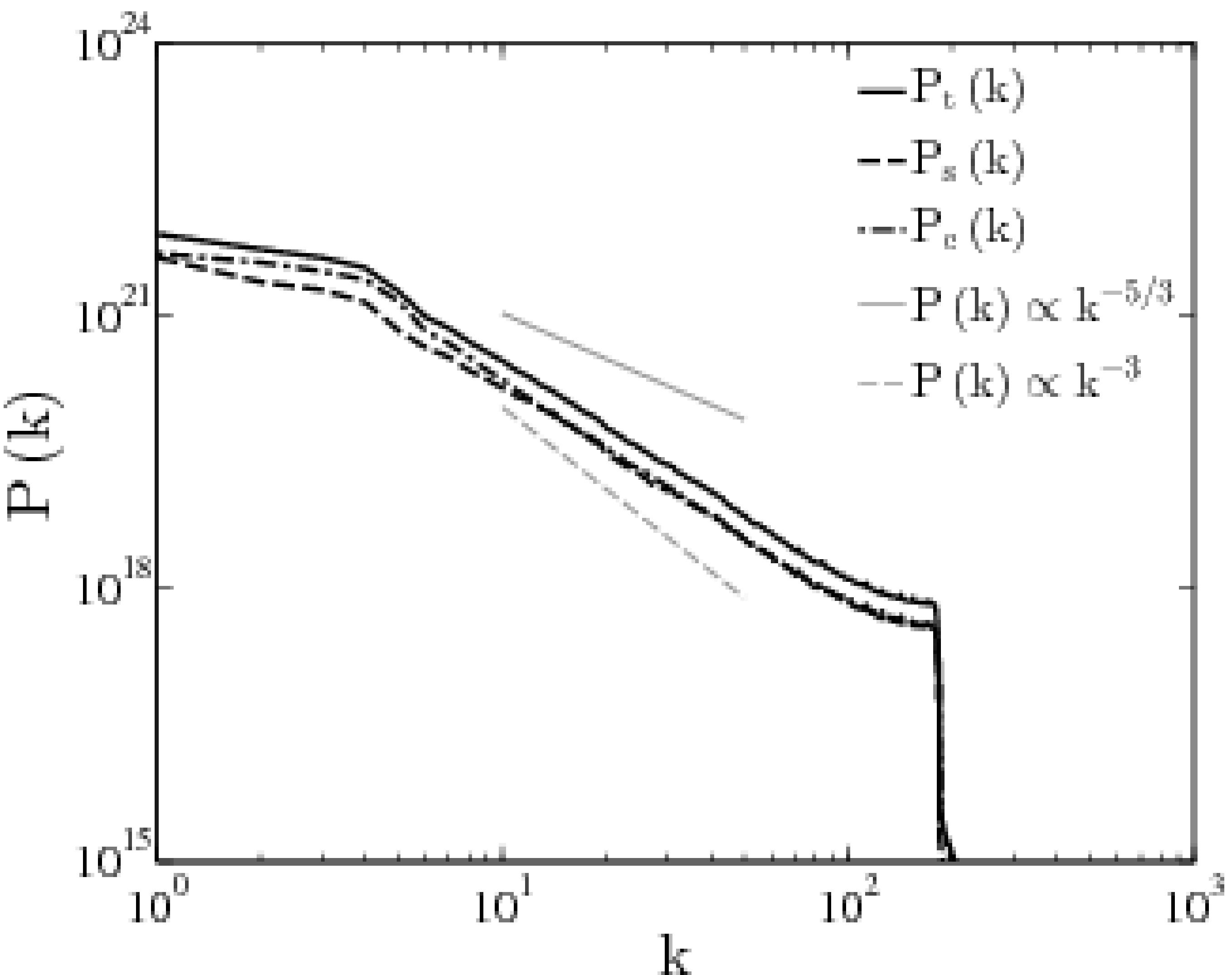}
			\put(20,15){$t=130$ ns}
		\end{overpic}
		~
		\begin{overpic}[width=5cm]{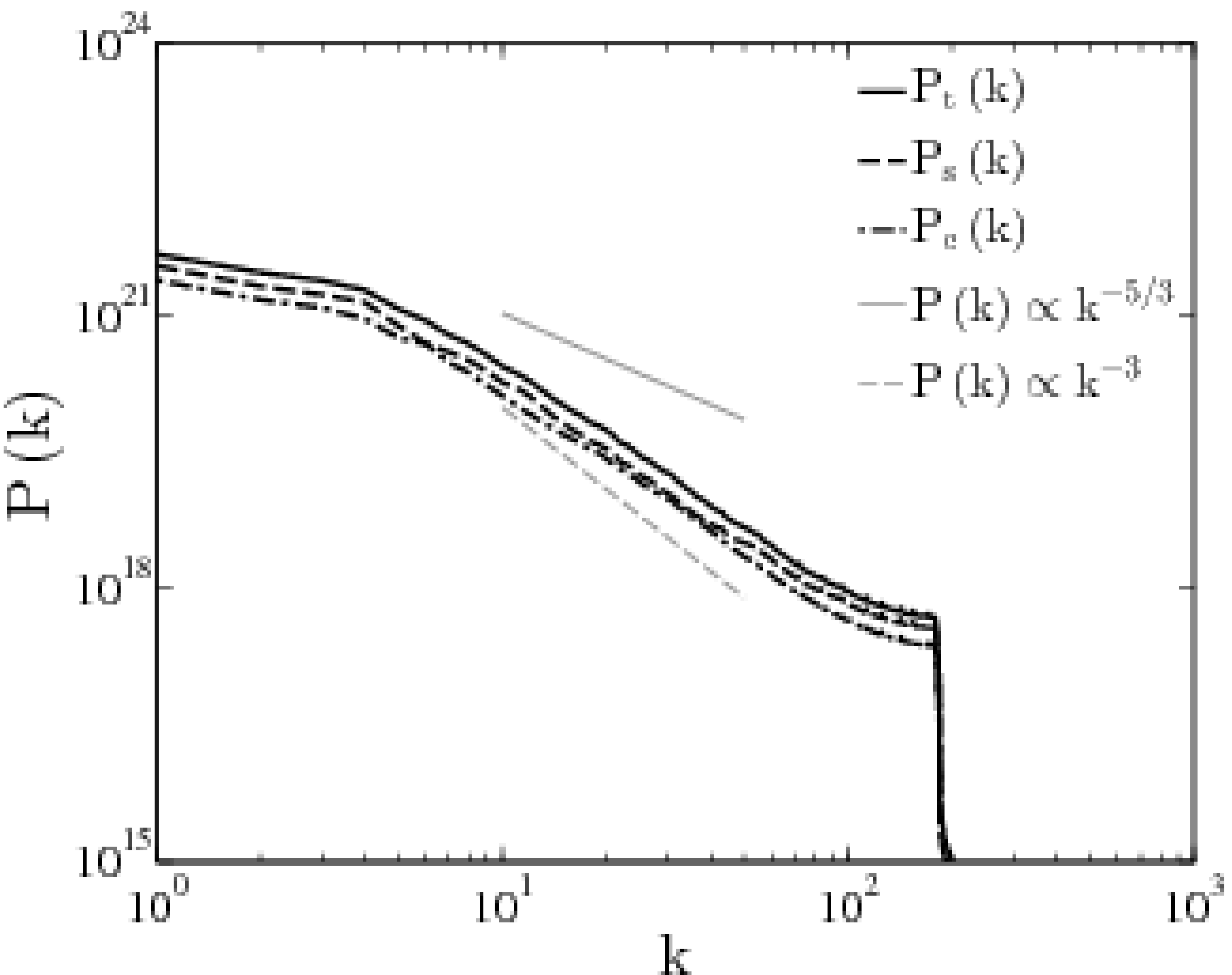}
			\put(20,15){$t=200$ ns}
		\end{overpic}
		~
		\begin{overpic}[width=5cm]{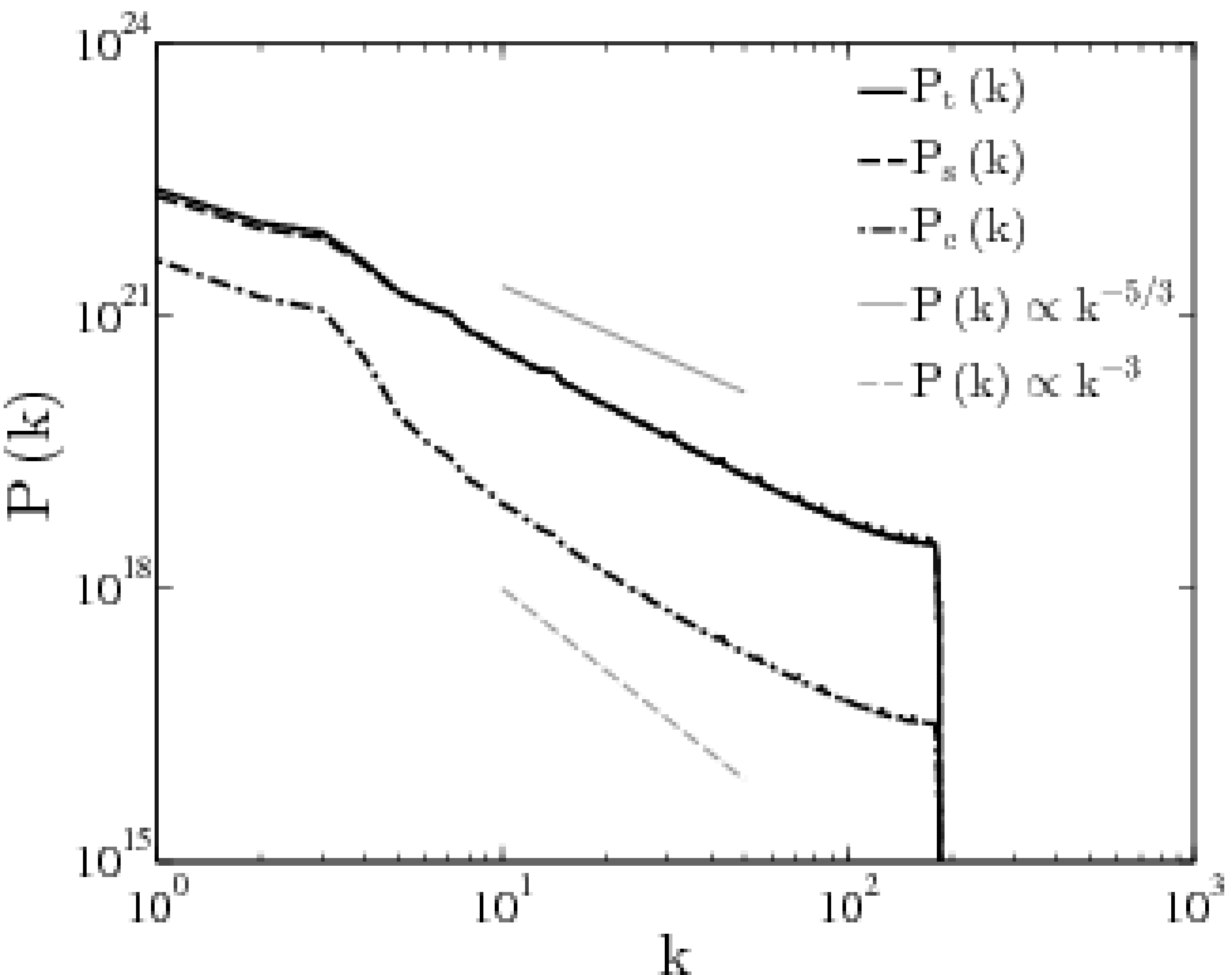}
			\put(20,15){$t=230$ ns}
		\end{overpic}
		
	\end{center}
\caption{Kinetic energy spectra (total, compressive, and solenoidal) at $t=130$~ns, $t=200$~ns, and $t=230$~ns for the self-generated magnetic field case. (a,b) During the driving phase the power spectra in the inertial range ($5 \lesssim k \lesssim 100$) scales as $P\propto k^{-2.3}$. This deviates from the traditional two-dimensional $k^{-3}$ behavior due to laser-driven stirring and not magnetic field effects. (c) After driving ceases, the solenoidal component becomes the dominant source of kinetic energy and scales as $P\propto k^{-2}$.  In the absence of inertial
forces, the dominant process is the interaction between vortical
cells, leading to a $k^{-2}$ dependence.}
\label{f:spectras}
\end{figure*}
\begin{figure}[h!]
	\begin{center}
		\includegraphics[width=8cm]{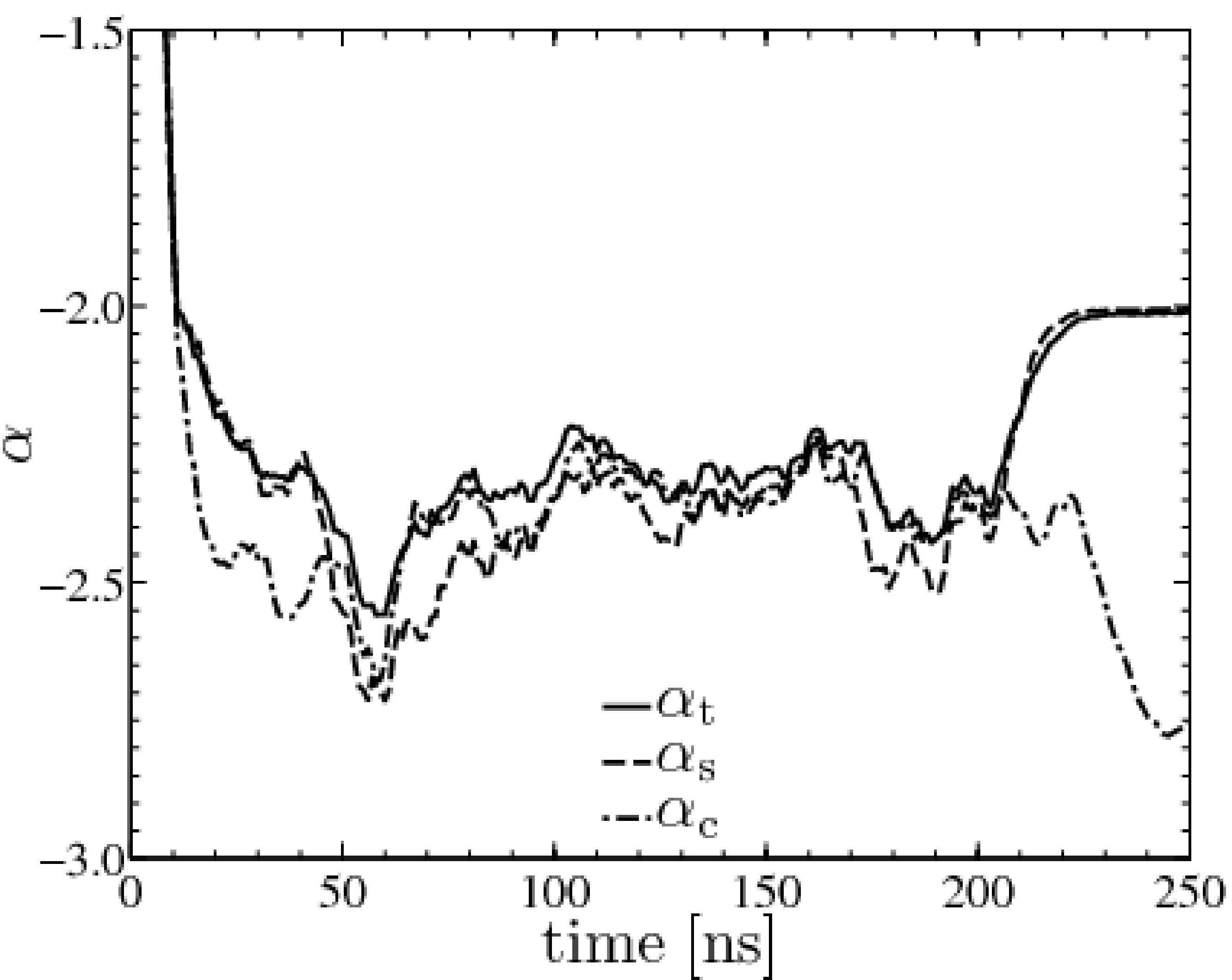}
	\end{center}
	\caption{Evolution of the velocity power spectra exponent for $10\le k\le 50$. The turbulent flow reaches a quasi-steady state near $t=75$~ns with $\alpha\approx-2.3$ for all components. This deviation from $\alpha=-3$ for two-dimensional turbulence is a consequence of the laser-driven stirring.}
	\label{f:powerfits}
\end{figure}
In order to compute the velocity power spectra on the mesh we consider a square area of 1000\micron by 1000\micron inside the region of interest. Our finest mesh (4\micron resolution) provides a uniform $512^2$ grid. Likewise, the 8\micron and 16\micron meshes produce $256^2$ and $128^2$ grids, respectively. From these grids we compute the numeric Fourier transform of $\bf{u}\left(\bf{x}\right)$, $\bf{\hat{u}}\left(\bf{k}\right)$, using the fast Fourier transform. We define the multi-frequency, total velocity power spectra as $P_t\left(\bf{k}\right)=\left|\bf{\hat{u}}\left(\bf{k}\right)\right|^2/2$. In order to reduce this to a function of a single, mean wavenumber $\left(k\right)$ we average $P_t\left(\bf{k}\right)$ over circular shells of unit thickness in $\bf{k}$-space. In addition to the spectrum for the total velocity field, we also examine the power spectra for the compressive and solenoidal components of the velocity field, $P_c\left(k\right)$ and $P_s\left(k\right)$.

Figure~\ref{f:spectras} shows estimated velocity power spectra at $t=130$~ns, $t=200$~ns, and $t=230$~ns for the 4\micron resolution, self-generating case. Figure~\ref{f:powerfits} shows the evolution of the power law exponent, $\alpha$, for the total, solenoidal, and compressive components. A least-squares fit over the range $10\le k \le 50$ is used to estimate $\alpha$. The profiles are smoothed over a $t=5$~ns window to remove high frequency oscillations and allowing differentiation of the curves.

In all snapshots the inverse energy cascade ($P\propto k^{-5/3}$) is visible up to the driving mode at $k=5$.

During the driving phase, the power spectra in the inertial range ($5 \lesssim k \lesssim 100$) scales as $P\propto k^{-2.3}$. This behavior is similar to the classical theory of magnetohydrodynamic turbulence, where two dimensional behavior obeys $P\propto k^{-7/3}$ \cite{kraichnan+65, kraichnan+80}. This relation is likely coincidental as the magnetic field in these two dimensional simulation is generated out-of-plane and acts only as an additional pressure term. The self-generated magnetic fields produce negligible pressures that are unable to drive material motion over the timescales considered. Indeed, these results agree with the same analysis run on the pure hydrodynamic case. As a result, the deviation from the expected $P\propto k^{-3}$ behavior is most likely due to the laser-driven stirring.

In the post-driving phase the behavior of the compressive and solenoidal power spectra diverge. The equilibration of pressure in the region of interest results in a substantially weaker contribution from compression effects. In the absence of inertial forces, the dominant process is the interaction between vortical cells. This makes the system appear diffusion-dominated rather than advection-dominated. This has certain consequences on the kinetic energy spectra, causing it to assume the form $P\propto k^{-2}$ (Fig.~\ref{f:powerfits}, $t > 225$~ns). This can be explained by transforming the diffusion operator into $k$-space, which gives a time-independent spectra with $k^{-2}$ dependence.

We note that it is typical to see the inertial region smoothly transition into numerical dissipation at high wave numbers. However, the inertial region flattens before abruptly dropping into the dissipation range (not shown is a smoothly decaying knee beginning after the sudden drop). We believe there are two possible causes for this bottleneck effect. 

A potential physical explanation is that the stirring mechanism limits the cascade of energy to smaller scales. In particular, shocks moving through the turbulent core will overrun and destroy small-scale features. As such, the driving process may impose a lower-limit on feature size resulting in a buildup of power near this lower bound. 

An alternative explanation for the bottleneck is that numerical effects halt the transfer of energy to smaller scales \cite{sytine+00}. As we were unable to successfully compute models using Riemann solvers other than HLLE, this hypothesis is difficult to either confirm or refute. 
\section{Discussion and conclusions}
\label{s:conc}

We have presented the results of a computational study of a high-energy density physics laser-driven experiment aimed at producing supersonic turbulence in plasma. The design included a target irradiated by sets of laser beams to provide plasma confinement and induce turbulence. To this end we computed a generic laser drive profile (LDP). During the evolution, the LDP has been mapped at select positions and times in such a way as to create turbulent conditions in the central region of the target. 

We found that:

\begin{itemize}

\item
{
	The turbulence Mach number reaches nominal, quasi-steady value of 0.2 throughout the driving phase for all cases. There is a minor downward trend as the material in the turbulent core is increasingly thermalized. Shortly after the driving phase, the turbulence Mach number rises due to the conversion of suppressed thermal energy into kinetic energy as the confining ram pressure is removed. In the post-driving phase the turbulence Mach number reaches steady values of about 0.25.
}
\item
{
	The magnetic fields produced for the out-of-plane \emph{a priori} field are on the order of megagauss. These fields correspond to $\beta\approx 10-100$. Amplification of the magnetic field due to driving results in a factor of 2 increase during the driving phase. In the post-driving phase the spatial distribution becomes uniform with a nominal value on the order of 100~kG.

	The self-generated magnetic fields obtain kilogauss strengths during the driving phase. These fields correspond to $\beta\approx 10^{4}-10^{5}$, indicating that the effects due the magnetic field on the hydrodynamic development of the system are minimal. 	
}
\item
{
	The distribution of material velocity obtains an isotropic distribution during the driving phase for the self-generated case.
}
\item
{
	The velocity power spectra show the expected inverse energy cascade and forward enstrophy cascade during the driving phase. The forward enstrophy cascade obeys $P\propto k^{-2.3}$ in the inertial range ($5\lesssim k \lesssim 100$). The deviation from the two-dimensional, hydrodynamic behavior of $P\propto k^{-3}$ is due to the laser-driven mechanism and not magnetic field effects.
}
\item
{
	The solenoidal and compressive kinetic energies are roughly in equipartition during active driving. In the post-driving phase, the solenoidal component contains the bulk of the kinetic energy (with a ratio on the order of 100:1).
}

\end{itemize}

We conclude that, in principle, one can produce a weakly compressible, quasi-steady state, turbulent plasma in laser driven experiments for as long as driving is provided. We note, with some disappointment, that the particular choice of parameters did not produce supersonic turbulence. We found this primarily due to the turbulent central region being filled with the ablating material. It would be interesting to consider a scenario in which the entire target is composed of a single material. In the case that the target is made out of the low density material used in this study, one could expect the turbulent Mach number to increase by a factor of $\sqrt{5}$ ($\approx$2.2).

Furthermore, it is conceivable that by adjusting the firing frequency of the laser, and the energy and pulse length of individual beams one can change the thermodynamic conditions in the turbulent core. This provides a way to control the sound speed, and therefore the turbulent Mach number opening a possibility of reaching the supersonic regime.

One aspect of the proposed design we did not discuss in this work in detail is experimental diagnostics. We defer the discussion of this crucial component of the experiment until more realistic computations are performed in three-dimensions.

\section{Acknowledgments}\label{s:ack}
TH and TP were supported in part by the DOE grant DE-FG52-09NA29548 and the NSF grant AST-1109113. The authors wish to thank Mike Grosskopf of the University of Michigan for his assistance with configuring the CRASH code. This research used resources of the National Energy Research Scientific Computing Center, which is supported by the Office of Science of the U.S.\ Department of Energy under Contract No.\ DE-AC02-05CH11231. The software used in this work was in part developed by the DOE Flash Center at the University of Chicago. 

\appendix
\section{Calculation of the Laser Drive Profile}
\label{s:crashlaser}

We compute the laser drive profile (LDP) using the CRASH code \cite{vanderholst+11}. In this configuration, a 2.5D cylinder of carbon foam is irradiated along the positive $z$-axis. The physical domain for the LDP simulations is $-120$\micron~$\le z\le 2000$\micron, and $0\le r\le 2000$\micron. The initial density is set to $\rho_{ambient}$ for $z<0$, and $\rho_{shell}$ for $z\ge0$. Initially, both layers are in pressure equilibrium at $1\times10^8$~Pa. These conditions match those of the target in the full Proteus simulation. The LDP simulations do not contain any magnetic field effects. 

The incident irradiation is defined by by a super-Gaussian laser beam of order $4.2$, with a standard deviation of $250$\micron~in the radial direction, and whose center is coincident to the axis of symmetry. The pulse is a constant $3.3\times10^{13}$~W for $t=1.8$~ns with a linear rise and decay time of $0.1$~ns. Laser energy deposition is accomplished using CRASH's geometric ray-tracing functionality with $400$ individual rays representing the beam. 

We performed two runs with uniform mesh resolutions of $8$\micron~and $4$\micron. These resolutions are at least a factor of two smaller than the Proteus mesh the LDP is mapped to. We determined there were minimal differences between the two resolutions. 

We choose the evolutionary snapshot to be used as the LDP by finding the moment when the rarefaction fan catches up to the shock front. The laser driven flow is thermally developed at this point, and the bulk of the energy transferred to the shell will be in the form of kinetic energy. This corresponds to the snapshot at $t=2.1$~ns in both the $8$\micron~and $4$\micron~simulations. This evolutionary time is after the CRASH laser drive turns off and we are not truncating any laser physics by choosing the $t=2.1$~ns snapshot.

\begin{figure*}[!htbp]
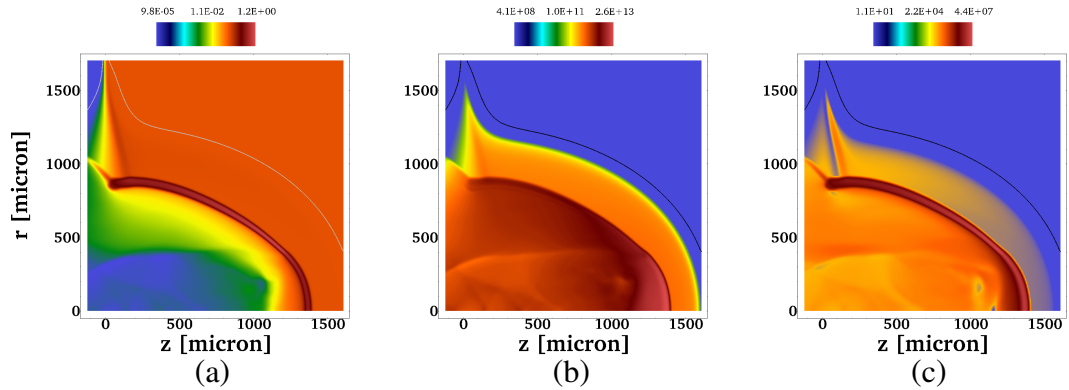

	\centering	


	\begin{overpic}[bb=0 0 4096 4096, clip, height=5cm]{f12_a_color.pdf}
		\put(50,-2){\large $\mathrm{\left(a\right)}$}
	\end{overpic}
	~
	\begin{overpic}[bb=410 0 4096 4096, clip, height=5cm]{f12_b_color.pdf}
		\put(40,-2){\large $\mathrm{\left(b\right)}$}
	\end{overpic}
	~
	\begin{overpic}[bb=410 0 4096 4096, clip, height=5cm]{f12_c_color.pdf}
		\put(40,-2){\large $\mathrm{\left(c\right)}$}
	\end{overpic}
		
	\caption{Pseudocolor plots of quantities for the prototypical laser drive calculated with CRASH. All plots are colored in log scale. The solid line which forms an envelope over the on all plots indicates our cutoff region for mapping ($p_\mathrm{cut}=3\times10^9$~Pa). (a) Density in g cm$^{-3}$. (b) Pressure in Pa. (c) Velocity magnitude in cm s$^{-1}$. }
	\label{f:crash}
\end{figure*}
Depicted in Fig.~\ref{f:crash} are the pseudocolor plots for density, thermal pressure, and velocity magnitude. The laser drive produces a parabolic structure moving predominately in the radial direction. By the time the flow becomes thermally developed the extent the blast wave reaches nearly $1500$\micron. The ablation near the beam leaves a low density, high pressure region behind the shock, implying that the bulk of the kinetic energy that will reach the core is compressed near the shock structure. In addition, the model shows a substantial preheat region ahead of the shock. 

The only qualitative difference between the $8$\micron~and $4$\micron~models is near the symmetry axis; the $4$\micron~model shows a slightly more bulbous structure. We do not think this will affect the generation of turbulence. Therefore, in the interest of memory constraints and mapping time, we choose to use the $8$\micron~model as the prototypical laser drive. 

\subsection{Mapping the laser drive profiles}

We interpret the configurations described in Sec.~\ref{s:expcond} as a set of cylinders (rectangles in two-dimensions) embedded into the shell of the target. The geometry of these cylinders are computed at code startup on each processor. Additionally, LDP hydrodynamic variables (obtained in Sec.~\ref{s:crashlaser}) are loaded onto each core. 

When it is decided that a laser should be fired in the Proteus simulation (described in Sec~\ref{s:expcond}), the cells owned by a processor are searched to determine if any reside inside, or are clipped by, the laser cylinder. If any do, the location of the cell in the cylinder's local coordinate system are determined. From this information, we determine where the
cell lies in the two-dimensional CRASH data set. In the case of two dimensions, this is intuitive as we are rotating and shifting a rectangle onto another rectangle. In three dimensions we only consider the axial and radial coordinates
in the cylindrical coordinate system, and discard the angular component. In our study this is reasonable, as we are mapping an axisymmetric data set. However, we note that mapping a three-dimensional data set into the cylinder (or any other geometric primitive) requires a physically meaningful definition of the rotation about the axis, complicating the modeling of the system.

After determining our current cell's position in the data set frame, we interpolate the data set quantities using bi-linear interpolation. In order to enhance the mapping obtained, we perform the interpolation on a uniform grid of 10 points per dimension, weighting by sub-cell volume. 

The quantities interpolated from LDP are forcibly written onto the mesh in the pre-computed location. This entails overwriting the values of density, pressure, and velocity components (transformed into the coordinates of the cylinder axis). We lessen obtrusiveness of mapping on the surroundings by only using the LDP data inside of a pressure cutoff of $p_{cut}=3\times10^8$ Pa. This cutoff is shown as the gray or black solid line moving through the domain in Fig.~\ref{f:crash}. Additionally, strong discontinuities near the edge of the mapped data can lead to failures of the Riemann solver. To circumvent this problem we smooth the fields with an arithmetic averaging filter of $5\times5$ cells near the edge. 

%
%
\bibliographystyle{model1a-num-names}
\bibliography{turb_hedp}
\end{document}